\documentclass[preprint2]{aastex}

\slugcomment{Not to appear in Nonlearned J., 45.}

\shorttitle{Multiple-Cores Systems in Cluster-Forming Regions}
\shortauthors{Saito et al.}

\begin{document}


\title{HIGH-RESOLUTION STUDIES OF THE MULTIPLE-CORE SYSTEMS TOWARD 
CLUSTER-FORMING REGIONS INCLUDING MASSIVE STARS}


\author{Hiro SAITO\altaffilmark{1}, Masao SAITO\altaffilmark{1}, 
Yoshinori YONEKURA\altaffilmark{2}, and Fumitaka NAKAMURA\altaffilmark{3}}
\email{saito@nro.nao.ac.jp}





\altaffiltext{1}{National Astronomical Observatory of Japan, Osawa 2-21-1, 
Mitaka, Tokyo 181-8588, Japan}
\altaffiltext{2}{Department of Physical Science, Osaka Prefecture University, 
Gakuen 1-1, Sakai, Osaka 599-8531, Japan}
\altaffiltext{3}{Faculty of Education and Human Sciences, Niigata University,
8050 Ikarashi2-no-cho, Niigata City 950-2181, Japan}


\begin{abstract}
We present the results of C$^{18}$O observations by the Nobeyama Millimeter 
Array toward dense clumps with radii of $\sim 0.3$ pc in six cluster-forming 
regions including massive (proto)stars. 
We identified 171 cores, whose radius, line width, and molecular mass range 
from 0.01 to 0.09 pc, 0.43 to 3.33 km s$^{-1}$, and 0.5 to 54.1 
$M_{\sun}$, respectively. 
Many cores with various line widths exist in one clump, and the index of 
the line width--radius relationship of the cores and the parental clump differs 
from core to core in the clump.
This indicates that the degree of dissipation of the turbulent motion varies 
for each core in one clump. 
Although the mass of the cores increases with the line width, most cores are 
gravitationally bound by the external pressure. 
In addition, the line width and the external pressure of the cores tend to 
decrease with the distance from the center of the clump, and these dependencies
may be caused by the inner H$_2$ density structure of the clump that affects 
the physical properties of the cores. 
Moreover, the number density of the cores and the number density of young 
(proto)stars have a similar relationship to the average H$_2$ density of 
the clumps. 
Thus, our findings suggest that the cluster is formed in the clump through 
the formation of such multiple cores, whose physical properties would have 
been strongly related to the H$_2$ density structure and the turbulent motion 
of the clump. 

\end{abstract}





\section{Introduction}

Most stars in the galactic disk are formed in dense molecular gas within 
giant molecular clouds (GMCs) as members of clusters (Lada \& Lada 2003). 
Thus, clusters must play a critical role in the origins of some of 
the most fundamental properties of the galactic stellar population. 
In addition, such clusters consist of stars of various masses, and 
the most massive stars exist at the centers of clusters 
(e.g., Raboud \& Mermilliod 1998). 
These features suggest that the physical environment in the cluster-forming 
region would vary from area to area because the mass of formed stars should 
be closely related to the physical environment.
However, clusters do not always include massive stars. Some clusters contain 
only intermediate-mass and/or low-mass stars in low-mass star-forming regions 
(e.g., Tachihara et al. 2002). 
Based on these features, it appears that although the conditions of massive 
star formation easily satisfy the physical conditions for cluster formation, 
the conditions of cluster formation do not necessarily provide the physical 
conditions for massive star formation.


The clusters appear to be continuously forming in the galactic disk and 
the direct study of their physical conditions and the processes leading to 
their formation is basically possible. 
However, discovering clusters at an early evolutionary stage via optical 
observations is difficult because such clusters are completely embedded in dense
gas and dust. Therefore, observation by infrared instruments is necessary to 
study cluster formation.
Recently, many young embedded clusters within molecular clouds have been 
discovered by infrared observations (e.g., Bica et al. 2003). On the basis of 
these observations, to investigate the physical conditions in the dense gas that
forms rich clusters, dense molecular gas surveys were carried out on such young
clusters that include massive stars, revealing that 
clusters are formed in gas systems with a size scale of a cluster 
($\sim 0.3$ pc) in GMCs (e.g., Lada 1992; Lada \& Lada 2003; Saito et al. 2007) 
and these systems are usually called ``clumps.'' 
However, a few very compact gas systems having a size scale of $\sim 0.03$ pc 
similar to typical (cold) cores in low-mass star-forming regions 
(e.g., Onishi et al. 2002; Umemoto et al. 2002) have been discovered with very 
high-angular-resolution observations using an interferometer on young clusters 
that include massive (proto)stars 
(e.g., Churchwell et al. 1992; Olmi et al. 1993). These systems have high 
temperatures ($> 100$ K), a high H$_2$ density ($10^6$--$10^7$ cm$^{-3}$), and 
a high luminosity ($> 10^4$ $L_{\sun}$). These results indicate that 
these systems called ``hot cores,'' would include massive protostar(s) or 
massive young star(s).

Generally, not a large number of stars form in a core even in cluster-forming 
regions (e.g., van der Tak et al. 2005). 
Hence, many cores must form in one clump to create a cluster with a large 
number of stars.
If this multiple-core system is formed by the fragmentation of the inner
structure of the clump, a close relationship should exist between the physical
parameters of the clump and the characteristics of the formed cluster.
Saito et al. (2007) carried out molecular line observations using the Nobeyama 
45-m telescope directed toward the clumps in several cluster-forming regions 
and suggested that a good relationship indeed exists between the average H$_2$ 
density of the clumps and the stellar number density of formed clusters in 
these clumps. In addition, they found a good correlation between the line width 
of the clumps and the maximum mass of formed stars.
Saito et al. (2006) investigated the cores in three cluster-forming regions
and found two types of cores in a single clump. The first type was a core with 
a large mass and a large turbulent motion, and the second type was a core with 
a small mass and a small turbulent motion. 
They found that massive (proto)stars are associated with the cores with 
a large mass and large turbulent motion.
This suggests a theoretical dependence between the mass of the star formed 
in a core and the turbulent motion in the core (e.g., McKee \& Tan 2003). 
According to the suggestion by McKee \& Tan (2003), 
the results of Saito et al. (2006) indicate that massive and low-mass stars 
can be formed from one clump.
Such observations suggest that to understand the mechanism of cluster 
formation, it is important to elucidate the relationships between the physical 
properties of cores and physical condition of the clump that includes the cores,
and between the mass of the star formed in a core and the physical condition of 
the core. If the cores in a clump are formed by gravitational instability as 
the theory suggests, a strong connection must exist between the spatial 
distribution of the cores in the clump and the inner structure of the clump.
However, no detailed relationships between the cores and the clumps have been 
demonstrated. In addition, although young clusters have a tendency to 
concentrate massive stars in the center, whether massive cores with large 
turbulent motion are formed in the center of the clump remains unclear.

Therefore, we must observe several clumps to obtain detailed physical parameters
of both clumps and cores and reveal the relationships between the cores and 
the clumps themselves. 
To detect the cores in the clump, we must uncover the internal motion 
and the distribution of the H$_2$ column density in the clump. 
Thus, we need to carry out molecular line observations suitable for revealing 
velocity structures of the dense gas in the clump.
Such observations are important to find a protocluster that consists of 
multiple cores --- a ``core cluster.''


We carried out C$^{18}$O observations toward nine IRAS point sources with 
young clusters containing massive stars. The C$^{18}$O line is a good 
candidate for tracing a large dynamic range in mass and column density in 
the dense gas because this line has a thin optical depth and a low critical 
density.
In general, the C$^{18}$O line cannot be used to probe high-density regions
($> 10^{4}$ cm$^{-3}$) in starless cores because CO molecules are depleted 
onto dust grains in such a cold region with a temperature lower than the CO 
sublimation temperature of $\sim 20$ K (e.g., Caselli et al. 1999; 
Bacmann et al. 2002).
However, previous CO observations have revealed that CO lines trace high-density
regions ($> 10^{4}$ cm$^{-3}$) well with high temperatures of $\gtrsim 20$ K 
(e.g., Momose et al. 1998; Sch\"{o}ier et al. 2004). 
Therefore, if the temperatures of high-density regions ($> 10^{4}$ cm$^{-3}$) 
are higher than $\sim 20$ K, C$^{18}$O lines can be used to probe 
these regions. 
Typically, the regions where massive stars are formed in GMCs maintain high 
temperatures of $\gtrsim 20$ K, and thus C$^{18}$O molecules would not be 
depleted in such regions.
In fact, the structures identified by the interferometric observations toward
massive star-forming regions have a high average density of 
$> 10^{5}$ cm$^{-3}$ (e.g., Hunter et al. 2004; Saito et al. 2006).
Therefore, C$^{18}$O observations can be used to reveal the distribution, mass, 
and kinetic motion of the dense gas in cluster-forming regions if this line is 
optically thin enough.


The identification of the star-forming and starless cores requires observation 
with high-angular resolution of several arc-seconds because massive star-forming
regions are generally located at a distance of several kpc. 
However, interferometric observation of C$^{18}$O cores in massive star-forming 
regions has rarely been performed.
To find dense cores in the clumps with cluster formation, we carried out 
C$^{18}$O molecular line observations with high-angular resolution toward 
nine cluster-forming regions using the Nobeyama Millimeter Array (NMA). 
The physical conditions of the clumps in these regions were already derived by 
Saito et al. (2007). Results of the interferometric observations toward three of
these nine regions were already reported in Saito et al. (2006). 
Here we present the results of the observations toward the other six regions.
We present our millimeter interferometric observations in $\S$ 2. 
In $\S$ 3, the observational results are presented for millimeter continuum and 
C$^{18}$O molecular line. We discuss the implications of our results for 
the physical condition of the dense cores in $\S$ 4. 
In $\S$ 5, we summarize the main results of our study.

\section{Observations}


\subsection{Samples}
To compare the physical conditions in the dense gas around young clusters 
containing young massive stars with the characteristics of the clusters,
we need to select targets at similar evolutionary stages and distances and with 
similar bolometric luminosities, because the observed physical parameters depend
on both the characteristics of formed stars (e.g., Saito et al. 2001) and 
the spatial resolutions. 
In addition, the evolution of an embedded stellar cluster is sensitively 
coupled to the evolution of its surrounding gas (e.g., Lada \& Lada 2003). 
Carpenter et al.\ (1990), Zinchenko et al.\ (1994, 1997, 1998), and 
Sridharan et al.\ (2002) carried out observations with low-angular resolution 
($\sim 60 \arcsec$) toward luminous IRAS point sources associated with 
H$_2$O/CH$_{3}$OH masers and/or compact H\,{\footnotesize II} regions in 
the northern sky and identified 93 massive clumps with a size of $> 1.0$ pc.
H$_2$O/CH$_{3}$OH masers and/or compact H\,{\footnotesize II} regions serve as 
the indicators of very young massive stellar objects in massive star-forming 
regions. 
We selected 14 IRAS point sources from these clumps on condition that 
the luminosity of the IRAS source is several times $10^{4}$ $L_{\sun}$, 
the distance is $\sim $2 -- 4 kpc, and the source is associated with 
both the indicators of very young massive stellar objects and NIR clusters.

We observed nine of these where the physical parameters of the dense clumps 
were already derived by single-dish observations with high-angular resolution 
of $\sim 15 \arcsec$ (Saito et al. 2007). 
In this paper, we report the observational results obtained toward six objects 
except for three objects that were already reported by Saito et al. (2006).
These IRAS sources are associated with the indicators of very young massive 
stellar objects:
(1) (U)C H\,{\footnotesize II} region(s) for IRAS\,05375+3540 
(e.g., Felli et al. 2006), IRAS\,06117+1350 (Turner \& Terzian 1985), 
IRAS\,19442+2427 (e.g., Barsony 1989), and IRAS\,19446+2505 
(e.g., G\'{o}mez et al. 1995). 
(2) very weak centimeter continuum sources for IRAS\,06056+2131, and 
IRAS\,06061+2151 (Kurtz et al. 1994). 
This indicates that these massive (proto)stars are at a similar evolutional
stage within several times 10$^{5}$ yr. 
Table 1 lists the position, distance, and luminosity of each source.

\subsection{Nobeyama Millimeter Array(NMA) Observations}
The observations were carried out from November 2002 through May 2005 with 
the six-element Nobeyama Millimeter Array (NMA) in the C$^{18}$O 
($J\!=\!1$--$0$; 109.782 GHz) line, and 98 GHz and 110 GHz continuum. 
The 8-h tracks were obtained in the most compact (designated as D) 
and intermediate (designated as C) configurations. 
The projected antenna baseline length ranged from 13 m to 163 m, yielding 
the typical resolution of $\sim 4 \arcsec$ when natural weighting is 
applied to all visibility data. The resolution of each observation is given 
in Table 2.  At 110 GHz, the primary beam of the 10-m telescope is about 
$62 \arcsec$.  

We used the SIS receivers, whose system noise temperatures in single sideband 
were $\sim 400$ K (including the atmosphere toward the zenith). 
Digital correlators UWBC and FX were configured for 1024 MHz and 32 MHz 
bandwidths with 128 and 1024 channels per baseline, respectively, yielding 
an equivalent resolution of 44 km s$^{-1}$ and 0.1 km s$^{-1}$ at 110 GHz, 
respectively. The UWBC can provide both the upper sideband (110 GHz) and 
the lower sideband data (98 GHz).

The gain calibrations were carried out for each upper and lower sideband 
every 10-15 min by observing quasars 0528+134, 0552+398, and 2023+336. 
The bandpass calibration was performed with 3C454.3 and 3C84. Uranus was 
observed at 3 mm to estimate the absolute flux scale. The uncertainties in
the flux scale are about 15\%.

The visibility data were calibrated using the NRO software package UVPROC2, 
and final maps were made using the CLEAN algorithm of the Astronomical 
Image Processing System (AIPS) by the National Radio Astronomy Observatory 
(NRAO\footnote{The NRAO is operated by the National Science Foundation by 
Associates Universities, Inc., under a cooperative agreement.}). 
The 3-mm continuum maps are constructed from line-free channels of 
a 44 km s$^{-1}$ wide window. The uncertainties of the position in the maps
are about $< 1 \arcsec$. The noise levels of the natural-weighted maps 
are summarized in Table 2.

\subsection{Definition of C$^{18}$O cores}
We defined a compact structure, a core, from the distribution of C$^{18}$O 
intensity to estimate physical properties of the dense condensations. 
We used the following procedures, similar to those adopted by Saito et al. 
(2007):

(1) To identify the velocity components in the observed fields, we checked 
    the C$^{18}$O molecular spectrum. If the C$^{18}$O spectrum has more than 
    two line features and the separations among their radial velocities are 
    larger than 0.5 km s$^{-1}$, each line feature was attributed to different 
    velocity components.

(2) We made a C$^{18}$O integrated intensity map for each velocity component.

(3) We searched for the i-th intensity peak (peak-i) outside previously defined
    cores (i.e., core[1], core[2], ..., core[i-1]) in the C$^{18}$O integrated
    intensity map.

(4) We drew a contour at the half-level of the intensity value at peak-i and 
    defined the region enclosed by the half-intensity contour as a core[i].

(5) In cases when sub-peaks exist within the core[i], if the intensity at 
    a sub-peak is larger than 6 $\sigma$ above the hollow between it and 
    peak-i, we excluded the region around the sub-peak by splitting 
    the core[i] along the hollow. 

(6) In cases when previously defined cores (i.e., core[1], ..., core[i-1])
    exist within the boundary of the core[i], if the intensity at peak-i is 
    larger than 6 $\sigma$ above the hollow between peak-i and the peaks of 
    the previously defined cores, we excluded the region around the peaks of
    the previously defined cores by splitting the core[i] along the hollow.  
    Otherwise, the core[i] was canceled.


(7) We searched for the next ([i+1]-th) intensity peak outside the cores.

(8) We repeated the procedures after procedure (4) until the value of 
    the intensity of the local peak went below the 6 $\sigma$ noise level. 

We identified 171 C$^{18}$O molecular cores using these procedures. 
The observed properties of the C$^{18}$O cores, such as the absolute peak 
temperature, $T^{\ast}_{\rm R}$, the radial velocity, $V_{\rm LSR}$, the FWHM line 
width, $\Delta V$, at the peak position, and the composite line width, 
$\Delta V_{\rm com}$, of the core are summarized in Table 3.
Here, the composite line width is the FWHM line width of the composite 
spectrum made by averaging all spectra within the core.
The typical uncertainties of the temperature, radial velocity, and line width 
in the Gaussian fits are 0.1 K, 0.1 km s$^{-1}$, and 0.1 km s$^{-1}$, 
respectively.

\subsection{Definition of the Millimeter Continuum Sources}
We next defined the millimeter continuum sources (MCSs) from the continuum 
images to estimate the physical properties of the very dense regions and to 
identify the positions of the massive (proto)stars.
Therefore, we identified MCSs except the known H\,{\footnotesize II} regions 
from the 98 GHz and 110 GHz continuum maps using procedures (3)--(8) adopted 
by the C$^{18}$O core definition.
We also stipulated that the MCSs were detected at both the 98 GHz and 
the 110 GHz continuum bands.
As a result, we identified 13 MCSs and detected six known 
(U)C H\,{\footnotesize II} regions. 
The observed parameters of these sources are summarized in Table 4.




\section{Results}
\subsection{Derivation of Physical Properties of the Cores}
The goal of this study was to clarify physical processes of both cluster
formation and massive star formation by investigating the indispensable 
physical characteristics of the dense gas that forms young clusters. 

In this paper, we defined a clump as a fundamental structure in which clusters
are formed and defined a core as a fundamental structure in which individual
stars in the cluster are formed.
Previous studies (e.g., Umemoto et al. 2002; Saito et al. 2006, 2007) indicated
that the clumps have a size scale of $\sim$ 0.3 pc and the cores have 
a size scale of $\sim 0.03$ pc.
The physical quantities of the dense gas in the cluster-forming regions, 
including massive star formations, are derived from the C$^{18}$O-line emission.

\subsubsection{C$^{18}$O line analysis}
Generally, the C$^{18}$O molecule is chemically stable and the optical depth 
of C$^{18}$O ($J\!=\!1$--$0$) is very thin. 
Thus, this line is very useful for estimating the H$_2$ column density.
Here, we calculate the C$^{18}$O optical depth and the H$_2$ column density 
under the local thermodynamic equilibrium (LTE) condition with the excitation 
temperature, $T_{\rm ex}$, estimated from the $^{12}$CO peak temperature of 
$T^{\ast}_{\rm R}$($^{12}$CO) = 31 -- 56 K at the position of each maser 
source  (Barsony 1989; Wouterloot \& Brand 1989; Zinchenko et al.\ 1998; 
White \& Fridlund 1992).
We assume that the excitation temperature of each core is the same as that at 
the position of the maser source in each region.
In these cases, the excitation temperatures, $T_{\rm ex}$, were estimated to 
be 35 -- 60 K. With these excitation temperatures, the C$^{18}$O optical depth, 
$\tau$(C$^{18}$O), and C$^{18}$O column density, $N$(C$^{18}$O), were estimated 
using
\begin{eqnarray}
\tau({\rm C^{18}O}) = -\ln\Biggm\{1 - \frac{T^{\ast}_{\rm R}({\rm
C^{18}O})}{5.27}\left[\frac{1}{\exp(5.27/T_{\rm
ex})-1} - 0.166\right]^{-1}\Biggm\},
\end{eqnarray}
and
\begin{eqnarray}
N({\rm C^{18}O}) = 2.42\times 10^{14}\times \Biggm\{\frac{\tau({\rm
C^{18}O})\Delta V({\rm km\ s^{-1}})T_{\rm ex}}
{\left[1 - \exp(-5.27/T_{\rm ex})\right]}\Biggm\} ({\rm cm^{-2}})
\end{eqnarray}
(e.g., Rohlfs \& Wilson 2004), where $T^{\ast}_{\rm R}$(C$^{18}$O) is 
the C$^{18}$O radiation temperature in kelvins. 
We calculated the H$_2$ column density, $N$(H$_2$), assuming that 
the C$^{18}$O abundance ratio is $1.7 \times 10^{-7}$ (Frerking et al.\ 1982). 
The C$^{18}$O optical depth and H$_2$ column density at 
the intensity peak of the cores range from 0.02 to 0.19 and from 
$0.8 \times 10^{22}$ to $2.4 \times 10^{23}$ cm$^{-2}$, respectively.
From these results, we found that the optical depth of the C$^{18}$O line 
is very thin.

\subsubsection{The Physical Parameters of the Cores}
First, we estimated the line width and the radius of the cores. 
We regard the line width of the composite line profile as the line width of 
the core. The line width of the cores ranges from 0.43 to 3.33 km s$^{-1}$ 
with a mean line width of 1.08 km s$^{-1}$.

We estimated the radius of the cores assuming that the cores are spherical. 
We defined the radius of the core, $R_{\rm core}$, as the deconvolved radius 
and calculated it as
\begin{eqnarray}
R_{\rm core}=\sqrt{\frac{S - {\rm beam\ area}}{\pi}},
\end{eqnarray}
where S and beam area are the area inside the core and the area of 
the synthesized beam, respectively. The radius ranges from $0.010$ to 
$0.090$ pc, with a mean radius of $0.035$ pc.
We selected the cores with the ratio of the deconvolved radius to the observed 
radius of $> 0.5$ to select the cores resolved by the present observations.
Note that we call the radius calculated from (S/$\pi$)$^{1/2}$ 
the ``observed radius.'' Hereafter, we use only the cores with high reliability
for discussions involving the radius.

Next, we estimated the LTE mass, $M_{\rm core}$, and the average H$_2$ density,
$n$(H$_2$)$_{\rm core}$, of the cores. 
The LTE mass of the cores was derived by integrating the H$_2$ column density 
over the core. We assumed a mean molecular weight per one H$_2$ molecule of 2.8 
by taking into account a relative helium abundance of 25\% by mass. 
The average H$_2$ density of the core is derived by assuming both a spherical 
shape and a uniform density and estimated by dividing the mass by a core 
volume, $(4/3)\, \pi R_{\rm core}\,^3$.
The LTE mass and the average H$_2$ density of the cores range from 0.5 to 54.1
$M_{\sun}$ and $0.6 \times 10^{5}$ to $2.1 \times 10^{6}$ cm$^{-3}$, 
respectively.

Finally, we estimated the virial mass, $M_{\rm vir}$, of the cores. 
The virial mass is one of the key quantities by which we can evaluate 
the equilibrium state of the cores. The virial mass of the core is derived 
using the following equation, assuming isothermal, spherical, and uniform 
density distribution with no external pressure or magnetic forces:

\begin{equation}
M_{\rm vir} = \frac{5\ R_{\rm core} (\Delta V_{\rm com})^{2}}{8\ G \ln 2},
\end{equation}
where $G$ is the gravitational constant. The virial mass of the cores ranges 
from 0.7 to 143.4 $M_{\sun}$. These physical properties are summarized in 
Table 5.

\subsection{Star-forming Cores and Starless Cores}
Here, we distinguish starless cores and star-forming cores to investigate 
the star-formation activity in C$^{18}$O cores in cluster-forming regions.
We compared the cores with near infrared sources (2MASS\footnote{The Two 
Micron All Sky Survey (2MASS) is a joint project of the University of 
Massachusetts and the Infrared Processing and Analysis Center/California 
Institute of Technology, funded by the National Aeronautics and Space 
Administration and the National Science Foundation 
(http://pegasus.phast.umass.edu).} sources), MCSs, and centimeter continuum 
sources (CCSs) as indicators of young stellar objects to reveal star formation 
in the cores. In particular, we selected the 2MASS sources detected at more than
or equal to two bands with $A_V > 10$ mag among the sources
associated with the clumps (Saito et al. 20007) because the sources embedded 
in the dense gas must have a large extinction such that the typical 
peak H$_2$ column density of the C$^{18}$O cores is more than 
$1 \times 10^{22}$ cm$^{-2}$. 
Note that the 2MASS sources identified by Saito et al. (2007) include young 
stellar objects (YSOs), which are the class I -- III type sources, and 
background sources.
Here, we define an association 
as when an intensity peak of one object is located inside the boundary of 
the other object.
Note that the cores associated with only MCSs are possibly very massive and 
dense starless cores.

First, we compared 13 MCSs with CCSs and 2MASS sources. 
Three of these are associated with both 2MASS sources and CCSs, and four of 
13 MCSs are associated only with 2MASS sources. 
Only one of 13 MCSs, MCS B for IRAS\,06056+2131, is associated only with CCSs 
and is seen as a dark lane in the $K_{\rm s}$-band image in Figures 2(b) and (c).
The flux density of the CCSs associated with MCSs corresponds to that of 
the H\,{\footnotesize II} region formed by a B2--3 star or of the ionized 
stellar wind formed by a B0--1 star (e.g., Panagia 1973; Felli \& Panagia 1981).
Therefore, the CCSs associated with MCSs would be caused by the activity of
massive YSOs.

We also derived spectral indices of MCSs from 98 GHz and 110 GHz data in 
Table 6. The spectral index of 8 MCSs with 2MASS sources and/or CCSs ranges 
from 2.2 to 4.6. 
This indicates that the millimeter continuum emission from these MCSs is 
subject to thermal dust radiation. 
Thus, the four MCSs with CCSs indicate the existence of young massive 
(proto)stars.
Three other MCSs only associated with 2MASS sources except for MCS A for 
IRAS\,19442+2427 have a strong intensity peak in millimeter continuum emission, 
and 2MASS sources associated with these MCSs have the brightness corresponding 
to an early B type or earlier star. 
Therefore these three MCSs would include early B type or earlier (proto)stars.
Indeed, two of these MCSs, MCS A for IRAS\,05375+3540 and MCS A for 
IRAS\,06117+1350, have a high luminosity of $\sim 1 \times 10^{3}$ $L_{\sun}$ 
(corresponding to a B3--4 star) and $\sim 7 \times 10^{4}$ $L_{\sun}$
(corresponding to a O9.5 star), respectively 
(Eiroa et al. 1994; Felli et al. 2004). 
Although the 2MASS source associated with MCS A for IRAS\,19442+2427 has
a brightness corresponding to an early B star, this MCS has no strong intensity
peak in millimeter continuum emission and the total integrated continuum 
emission from the MCS is very weak.
In addition, although this MCS has a large H$_2$ column density 
($\sim 10^{23}$ cm$^{-2}$), the associated bright 2MASS source has no large 
extinction ($A_V < 30$)
Thus, this 2MASS source would be located in front of the MCS.

However, four of five MCSs with neither 2MASS sources nor CCSs, MCSs A, B, and C
for IRAS\,19446+2505 and MCS C for IRAS\,19442+2427, have a flux density of 
110 GHz similar to or smaller than that of 98 GHz, and the spectral indices of 
these four MCSs are to be in the range between $-2.9$ and $+0.6$ in Table 6.
In particular, three MCSs for IRAS\,19446+2505 are located at the edge of 
the optical H\,{\footnotesize II} region, S88\,B. Thus, we regard these four 
MCSs as consisting of only free-free emission. 
Note that MCS B for IRAS\,19446+2505 with a large negative index of $-2.9$ is
more extended diffusely than the other MCS. Generally, it is probable that
the missing flux of the 110 GHz observation is larger than that of the 98 GHz
observation for a diffuse extended source because the minimum UV length
of the 110 GHz observation is longer than that of the 98 GHz observation.
Therefore, the MCS B would have such a negative index.
Thus, we regard this MCS as a source consisting of only free-free emission.
The other MCS with neither 2MASS sources nor CCSs, MCS B for IRAS\,19442+2427, 
has a strong intensity peak and a large spectral index of $\sim 3.7$, which is 
similar to that of the other massive-star-forming MCSs. 
Hence, in this paper, we regard this MCS as a source with massive protostars.

Next, we compared 171 C$^{18}$O cores with MCSs, 2MASS sources, and CCSs.
Eight of the cores are associated with both MCSs and 2MASS sources and two of 
171 cores are associated only with MCSs. In addition, 19 of 171 cores are 
associated only with 2MASS sources. From this result, all MCSs except for 
four MCSs classified as F.F. (see Table 6) are associated with C$^{18}$O cores.
However, cores N and R for IRAS\,19442+2427 are associated with same 2MASS 
source and same MCS, MCS A. The intensity distribution of the MCS corresponds
to the part with the high H$_2$ column density of these cores in Figures 5(b) 
and (d). The MCS would be associated with high-density part of these cores.
Therefore, cores N and R for IRAS\,19442+2427 with MCS A, which consist of 
cold dust emission, would be starless cores rather than cores with massive
YSOs. 
From these results, we regard 27 cores associated with 2MASS sources and/or 
MCSs except for cores N and R for IRAS\,19442+2427 as star-forming cores. 
In particular, we regard eight star-forming cores associated with MCSs that 
include massive protostars as massive-star-forming cores.

Therefore, we identified 27 star-forming cores, including eight 
massive-star-forming cores. These results are summarized in Table 5.
Note that the NIR source associated with these cores is not necessarily 
a (proto)star itself, and the possibility exists that this NIR emission is 
a reflection nebula of one or more central heating source(s).
In addition, it may be probable that some of the cores cataloged as starless 
are associated with mid-IR sources.

\subsection{Mass of the Continuum Sources}
Generally, the 98 and 110 GHz continuum emissions consist of free-free 
emission and/or thermal dust emission. First, we derived the flux of thermal 
dust component in MCSs with CCSs. The flux density of CCSs are much smaller 
than the total flux of MCSs at the millimeter wavelength as shown in Table 4. 
The spectral indices at 100 GHz of UC H\,{\footnotesize II} regions and ionized 
winds are usually $\sim -0.1$ and $\sim 0.5$ (e.g., Panagia \& Felli 1975), 
respectively. 
We assumed that these CCSs consist of the free-free emission with a spectral 
index of 0 
from an optically thin H\,{\footnotesize II} region or ionized winds,
and we estimated the flux of thermal dust component in MCSs by subtracting
the contribution of the free-free emission from the total flux of MCSs.

Next, we estimate the physical properties of the MCSs. The deconvolved radii 
of the MCSs, $R_{\rm d}$, are calculated using equation (3). 
The estimated radius of MCSs ranges from 0.018 to 0.110 pc. In particular, 
the radius of the MCSs associated with C$^{18}$O cores ranges from 0.018 to 
0.055 pc and is smaller than that of the associated cores
 ($R_{\rm d}$/$R_{\rm core} \sim 0.8$).
In addition, assuming an optically thin dust thermal emission at 100 GHz, we 
estimated the mass of MCSs using
\begin{equation}
M_{\rm d} = \frac{F_{\nu}\,D^{2}}{\kappa_{\nu}B(\nu,T_{\rm d})},
\end{equation}
where  $M_{\rm d}$ is the gas and dust mass of an MCS, $F_{\nu}$ is the total 
flux of the MCS, $B$($\nu,T_{\rm d}$) is the Planck function with the assumed 
dust temperature, $T_{\rm d}$, and 
$\kappa_{\nu} = \kappa_{\rm 230 GHz}(\frac{\nu}{\rm 230 GHz})^{\beta}$
is the dust opacity per gram with 
$\kappa_{\rm 230 GHz} = 0.005$ cm$^{2}$ g$^{-1}$ (Preibisch et al.\ 1993), which
assumes a standard gas-to-dust mass ratio of 100. 
Here, we assume that the dust opacity index, $\beta$, of the dust in MCSs is 
1.5 and that the dust temperature is equal to the temperature of the gas.
First, we need to determine the temperatures of MCSs to estimate the mass of 
MCSs. The temperature of MCSs A and B for IRAS\,05375+3540 were 
estimated by Felli et al. (2004) using CH$_3$CN lines, and we use 
this temperature as the dust temperature of the MCS.
Next, because no observation has been carried out toward the other MCSs using 
molecular lines, which trace the gas temperatures such as CH$_{3}$CN and 
NH$_{3}$, we estimate the temperature of MCSs with massive star formation to be 
$\sim 54$ -- $77$ K by applying the Stefan--Boltzmann law using the luminosity 
of (proto)stars associated with the MCSs.
If we estimate the temperature of MCSs A and B for IRAS\,05375+3540 using 
the same method, the temperature of $\sim 30$ K is obtained, which is similar 
to the temperature derived by molecular line described above.
Finally, concerning MCS A for IRAS\,19442+2427 without massive star formation, 
we use the excitation temperature estimated from the $^{12}$CO peak temperature 
as the dust temperature.
We estimate the mass of MCSs using these dust temperatures and the mass of 
the MCSs ranges from $\sim 2.7$ to $111$ $M_{\sun}$. We estimate the average 
H$_2$ density of the MCSs using the radius and the mass of the MCSs, and 
the average H$_2$ density ranges from 1.0 $\times 10^{6}$ to 
8.1 $\times 10^{6}$ cm$^{-3}$.

\subsection{Individual Regions}
The integrated intensity maps of C$^{18}$O as well as various kinds of objects
related to star formation are given in Figures 1 -- 6. For each region,
the distribution of C$^{18}$O clumps, which were taken from Saito et al. (2007),
IRAS sources, H$_2$O masers, H\,{\footnotesize II} regions, and the primary 
beam area of the present NMA observation are given in Figure (a).
In addition, the physical properties of the clumps are summarized in Table 7.
The integrated intensity distribution of C$^{18}$O in the clumps is compared 
with the 98 GHz or 110 GHz continuum image in Figure (b). 
In Figure (c), the C$^{18}$O distribution is overlaid on 
the 2MASS $K_{\rm s}$-band image. 
The distribution of C$^{18}$O cores, CCSs, and MCSs are given in Figure (d).
The strong peak in the continuum image indicate massive (proto)stars embedded 
in dense gas and the $K_{\rm s}$-band sources associated with cores indicate
young stars or protostars. 
In addition, although each missing flux in the NMA primary beam area is roughly 
estimated to be $\sim 70$\% using the single dish observation data 
(Saito et al. 2007), the missing flux toward the present C$^{18}$O cores is 
small to be $\sim 30$\%.
This suggests that the missing flux of the present observations would be caused
by the structure more extended than the size of the one measured with 
the present observations. 
As the interferometer filters out the extended emission and C$^{18}$O 
observations can detect the dense gas, the present observations would accurately
detect compact cores in a clump.
In the following sections, we present descriptions of the individual regions. 

\subsubsection{IRAS\,05375+3540 (GL\,5162, S235A/B)}
In Figure 1, we identified 21 C$^{18}$O cores and  two MCSs in three clumps,
clumps B, C, and D identified by Saito et al. (2007).
In addition, we detected free-free emission from a part of 
the H\,{\footnotesize II} region, S235 A (e.g., Felli et al. 2006), by 
the present continuum observation. 
C$^{18}$O core J is associated with MCS A and a bright 2MASS source. 
Although this 2MASS source is associated with the intensity peak of MCS A, 
the intensity peak of core J has an offset by $\sim 4 \arcsec$ from 
this 2MASS source. C$^{18}$O core H is associated with MCS B, a bright 2MASS 
source, and a CCS (Felli et al. 2006).

\subsubsection{IRAS 06056+2131 (GL\,6336s)}
In Figure 2, we identified 19 C$^{18}$O cores and two MCSs in clump A identified
by Saito et al. (2007).
C$^{18}$O core E is associated with MCS A and a bright 2MASS source.
Although the intensity peak of MCS A is associated with this 2MASS source,
the intensity peak of core E has an offset by $\sim 6 \arcsec$ from
this 2MASS source.
C$^{18}$O core J is associated with MCS B and two CCSs and the intensity peak 
of this core is associated with a dark lane in the $K_{\rm s}$-band image.

\subsubsection{IRAS 06061+2151 (GL\,5182)}
In Figure 3, we identified 24 C$^{18}$O cores and two MCSs in clump B identified
by Saito et al. (2007).
C$^{18}$O core O is associated with MCS B, two CCSs, and two bright 2MASS
sources. The intensity peak of this core is associated with one of these CCS 
and is located between these 2MASS sources.
C$^{18}$O core P is associated with MCS A, a CCS, and a bright 2MASS source.
The intensity peak of core P has an offset by $\sim 2 \arcsec$ from this
CCS. In addition, this CCS has a larger flux density ($\sim 3.4$ mJy) than 
that ($\sim 0.6$ mJy) of the CCSs associated with core O.

\subsubsection{IRAS 06117+1350 (GL\,902; S269)}
In Figure 4, we identified 28 C$^{18}$O cores and one MCS in two clumps,
clumps C and D identified by Saito et al. (2007).
We also detected the free-free emission from a part of 
the H\,{\footnotesize II} region, S269 (e.g., Turner \& Terzian 1985), by 
the present continuum observation.
The intensity peak of C$^{18}$O core M is located at both an intensity peak of 
MCS A and the position of a bright 2MASS source, IRS\,2w. According to 
the luminosity of IRS\,2w ($\sim 7 \times 10^{4} L_{\sun}$), C$^{18}$O core M 
may have an embedded massive (proto)star.

\subsubsection{IRAS 19442+2427 (GL\,2454; S87)}
In Figure 5, we identified 38 C$^{18}$O cores, three MCSs in three clumps, 
clumps A, B, and C identified by Saito et al. (2007).
In addition, we detected the free-free emission from both the core and 
the extended structure of H\,{\footnotesize II} region S87 (e.g., Barsony 1989).
Two velocity components, 21 km s$^{-1}$ and 24 km s$^{-1}$ components, exist 
in this region, and it has been suggested that the cluster, GL\,2454, was 
formed by the cloud--cloud collision of these components (Saito et al. 2007). 
C$^{18}$O cores, cores R, U, and AC, around the core structure of 
H\,{\footnotesize II} region S87 have a large mass ($\sim 10$ $M_{\sun}$) and 
a large line width ($\sim 1.5$ km s$^{-1}$).
C$^{18}$O core AH is associated with MCS B, which has a strong intensity peak
and a large spectral index, although this core is not associated with bright
2MASS sources. This indicates that the core could harbor an embedded massive 
YSO(s). In addition, the position of the intensity peak of this core in  
20.7 km s$^{-1}$ component map has changed from that in 21.4 km s$^{-1}$ 
component map in Figure 5(d), and MCS B is located between the intensity peak
in each component map. This indicates that core AH has a velocity gradient
and could have a disklike system embedded in it.
We suggest that this embedded massive YSO(s) in MCS B (see $\S$ 3.2) would be 
a massive protostar.

\subsubsection{IRAS 19446+2505 (GL\,2455; S88B)}
In Figure 6, we identified 41 C$^{18}$O cores, three MCSs in two clumps, clumps
B and C identified by Saito et al. (2007).
In addition, we detected millimeter continuum emission, which corresponds to 
two compact H\,{\footnotesize II} regions S88B1 and S88B2 
(e.g., G\'{o}mez et al. 1995).
The distribution of the C$^{18}$O emission is consistent with that of the dark
region in the $K_{\rm s}$ band.
C$^{18}$O cores, cores Q and W, around the compact H\,{\footnotesize II} 
region S88B1 have a large mass ($\sim 20 M_{\sun}$) and a large line width 
($\sim 2.0$ km s$^{-1}$).
Three MCSs exist at the boundary between the dense molecular gas and 
the optical H\,{\footnotesize II} region S88B. This indicates that 
the surface of the dense molecular gas would be ionized by the UV radiation 
from the massive stars in GL\,2455.

\section{Discussion}
\subsection{Physical Conditions of C$^{18}$O Cores}
Saito et al. (2007) identified 39 clumps with a size scale of $\sim 0.3$ pc 
in cluster-forming regions including massive star formation, and discussed 
their physical conditions by analyzing the correlations between the physical 
quantities of the clumps.
They found that the average H$_2$ density of the clumps increases with 
the line width and the clumps are self-gravitationally bound because
they have a virial mass comparable to a gas mass.
Here, we extend their analysis to the inner structure of the clumps (i.e.,
dense cores) and attempt to clarify the physical condition of the dense
cores that are the direct site of each star formation in 
the cluster-forming region.

First, we analyzed the relationships among the physical parameters of 
the cores. 
We used 171 C$^{18}$O cores identified in this study and 28 C$^{18}$O 
cores identified by Saito et al. (2006). 
The cores identified by Saito et al. (2006) were classified into
three massive-star-forming cores, seven star-forming cores, and 
18 starless cores using the classification adopted in this study.
These cores exist in 18 dense clumps identified by Saito et al. (2007).
Next we discuss the physical conditions of the dense cores inside the clumps
in cluster-forming regions including massive star formation.

\subsubsection{Line Width -- Radius Relationship}
The line width of molecular line emissions provides us with internal kinetic 
properties such as thermal motion and turbulence.
Several studies have suggested that a correlation exists between the line 
width and the radius of the core/cloud in the form of a single power-law
relationship $\Delta V \sim R^{\alpha}$; $\Delta V$ and $R$ are 
a characteristic line width and the radius of the core/cloud 
(e.g., Larson 1981). In addition, the index $\alpha$ has recently been shown to 
depend on the environment (e.g., Caselli \& Myers 1995; Saito et al. 2006). 

The left panel of Figure 7 shows the relationship between the line width and 
the radius of the 181 cores with high reliability (see $\S$ 3.1.2) identified
by this work and Saito et al. (2006) and 18 clumps identified by 
Saito et al. (2007).
Note the other 18 cores are those with a large error in the radius 
in $\S$ 3.1.2.
Although the line width of the clumps is in a narrow range of 
1.5 to 3.2 km s$^{-1}$, the line width of the cores is in a much wider 
range of 0.43 to 3.33 km s$^{-1}$.
In contrast, the range of the radius of the cores is relatively small.
Such features are seen even in the individual regions in the right panels of 
Figure 7.

We investigated the index $\alpha$ of the relationship between the line width 
and the radius of the core and the parental clumps. We estimated 
this relationship, $\Delta V_{\rm core}$ = $\Delta V_{\rm clump} \times$ 
($R_{\rm core}/R_{\rm clump}$)$^{\alpha}$, for all cores. 
The index $\alpha$ of this relationship ranges from 0.06 to 0.89 and 
the long-- and the short--dashed lines in Figure 7 indicate 
the line width--radius relationships with the minimum and maximum index 
$\alpha$, respectively,
using a mean line width and a mean radius of the clumps in the plots.
Our results show that the relationship between the line width and the radius
does not follow a single power-law relation. This trend agrees with the result 
of Saito et al. (2006), who used a smaller number of cores and clumps.
This indicates that the degree of the dissipation of the turbulent motion 
varies even within a single clump.
As discussed below in $\S$ 4.1.3, this may be due to the difference in 
the initial H$_2$ density or the external pressure of the core.
We postulate that the difference in the dissipation of the turbulent motion
depends on the region where the core is formed in the clump, considering 
that the clump generally has a density structure.

\subsubsection{Stability of Cores in the Dense Clump}
Generally, a core has to be gravitationally bound to result in star formation. 
Thus, a core with a large line width needs to have a large mass 
or high external pressure for star formation to occur.
Here, we check the virial condition of the cores to investigate the stability of
the system. Figure 8 shows correlations among the LTE mass, $M_{\rm LTE}$, and 
the virial mass, $M_{\rm vir}$, for 181 cores.
Most of the cores have a larger virial mass than the LTE mass, and no clear 
difference is seen between star-forming cores and starless cores. 
The cores have a virial ratio (the ratio of the virial mass to the LTE mass) of 
$\sim 2$, although several massive-star-forming cores have somewhat larger 
virial ratios of $\sim 4$.

Generally, more extended structures with star formation (such as clumps) have
a similar virial mass to a LTE mass (Fontani et al. 2002; Saito et al. 2007).
The main reason for the difference in the virial ratio between the clumps and
the cores is the elapsed time after the structures were formed. The virial mass
of the clump just after formation would be also larger than the LTE mass and
the clump would be gravitationally bound by the external pressure. 
The virial mass of the clump will later become similar to the gas mass because 
the density structure and the internal kinetic motion of the clump change and 
cluster formation is expected to occur in the clump. 
Therefore, almost all of the clump would be virialized as far as we observe
the clumps with cluster formation. 
However, many cores would also exist just after formation in the clump. 
Thus, many ``young cores,'' which have a larger virial mass than 
the gas mass, could be detected by the present observation.

In addition, the virial mass of the cores, except for 
massive-star-forming cores,
is approximately fitted by a single power-law function by
\begin{equation}
\log(M_{\rm vir}) = (0.17 \pm 0.03) + (1.11 \pm 0.04) \log(M_{\rm LTE}),
\, C.C. = 0.87,
\end{equation}
where C.C. is a correlation coefficient.
The coefficient of this relationship is close to unity, indicating that 
in the virial equation (see below), the difference in the gravitational and 
kinetic energy terms is almost proportional to the gravitational energy terms. 
To strike a good balance between this difference and the external pressure and 
for more massive cores to achieve the equilibrium state, 
the external pressure on the surface of the cores must be larger.

Most of the cores require external pressure to maintain the structures.
Here, we estimate the external pressure required to bind the present cores 
gravitationally. The virial equation is expressed by neglecting the magnetic 
fields as follows:
\begin{equation}
F = 2U + \Omega - 4 \pi R_{\rm core}^{3} P_{\rm ex}; 
(U = \frac{1}{2} M \sigma^{2}, 
\Omega = -\frac{3}{5} G \frac{M^{2}}{R_{\rm core}}),
\end{equation}
where $U$ and $\Omega$ are the kinetic and gravitational energy of the core,
respectively, $P_{\rm ex}$ is the external pressure on the core surface, and 
$G$ and $\sigma$ are the gravitational constant and the velocity dispersion, 
$\Delta V_{\rm com}$/($2 \sqrt{(2 \ln 2) /3}$). 
The external pressure required to bind the core gravitationally, $P_{\rm R}/k$,
can be estimated using these equations with $F$=$0$. The required pressure of 
the cores with a virial ratio of $> 1.0$ ranges from $1.3 \times 10^{5}$ to 
$4.3 \times 10^{8}$ K cm$^{-1}$.

Next, we investigated the external pressure on the core surface. 
Since the external pressure is difficult to estimate, we used the average 
pressure in a clump (average clump pressure) 
$\overline{P_{\rm CL}}/k$ instead, although these values generally differ from
each other. The average clump pressure was estimated using both the line width, 
$\Delta V_{\rm clump}$, and the average H$_2$ density, 
$\overline{{n({\rm H_2})}_{\rm clump}}$, of the clump. 
We estimated the average clump pressure using the parameters of the clumps 
identified by Saito et al. (2007) in Table 7 and found that the average clump 
pressure ranges from $1.1 \times 10^{7}$ to $1.9 \times 10^{8}$ 
K cm$^{-1}$.
Figure 9 shows the required pressure of cores plotted against the projected 
distance from the center of the clump to the core (core distance), $R_{\rm cen}$.
Here, the required pressure and the core distance are normalized to the average 
clump pressure and the radius of the clump, $R_{\rm CL}$, respectively. 
We found that the required pressure for most of the cores, except for 
massive-star-forming cores, is smaller than the average clump pressure.
Therefore, our findings suggest that most of the cores are likely to be 
gravitationally bound by the external pressure.
 
Finally, we investigated how the required pressure of cores changes with 
the position in a clump. 
In Figure 9, the cores with a large $P_{\rm R}$/$\overline{P_{\rm CL}}$ 
are concentrated in the center of the clump. 
Therefore, the nearer the cores are to the center of the clump, the higher
the external pressure on the core surface is expected to be.
When the H$_2$ density structure of the clump follows 
$n$(H$_2$)$_{\rm clump} \propto \,R_{\rm cen}\,^{-\gamma}$, the pressure in 
the clump (clump pressure), $P_{\rm CL}$, is estimated as a function of 
$R_{\rm cen}$ by $P_{\rm CL}$/$\overline{P_{\rm CL}}$ = 
($1 - \gamma$/3)$\,(R_{\rm cen}$/$R_{\rm CL}$)$^{-\gamma}$.
Here, we assumed that the turbulent velocity is constant within the clump.
For comparison, the distributions of the clump pressure for $\gamma$ = 
1.0, 1.5, and 2.0 are indicated in Figure 9 by the dotted, solid, and dashed 
lines, respectively.
These lines are similar to the upper limit of 
the $P_{\rm R}$/$\overline{P_{\rm CL}}$--$R_{\rm cen}$/$R_{\rm CL}$ plot, 
which indicates that almost all cores would be gravitationally bound by 
the external pressure if the H$_2$ density structure in the clump has a relation
 of $n$(H$_2$)$_{\rm clump} \propto \,R_{\rm cen}\,^{-\gamma}$.
In addition, these results suggest that the external pressure of the core 
surface would depend on the H$_2$ density structure in the clump.

\subsubsection{Mass -- Line Width Relationship}
Here, we investigated the relationship between the LTE mass and the line width 
of the core. Figure 10 shows that the LTE mass of the cores increases in 
proportion to the square of the line width, although the dispersion of 
the plot is large.  
This tendency agrees with the result of $\S$ 4.1.2: 
from equations (4) and (6), we obtain the relationship $M_{\rm LTE} \propto 
(\Delta V_{\rm core})^2$ considering that the radius of the cores is almost 
constant.
From this tendency, dense gas with a large kinetic motion in a clump seems to 
be necessary to form massive cores because the LTE mass of the cores increases 
with the line width.

Figure 10 also indicates that no clear difference exists between star-forming 
and starless cores, although several massive-star-forming cores (black circles) 
have somewhat larger line widths than the other cores with a similar mass.
For example, the average line widths of massive-star-forming cores and 
the starless cores (massive starless cores) with a mass $> 16$ $M_{\sun}$ are 
calculated as 2.8 km s$^{-1}$ and 2.0 km s$^{-1}$, respectively. 
The main cause of the difference in the line width is supposed to be the effect 
of star formation. If massive stars in massive-star-forming cores have formed 
from massive starless cores with a similar mass of $16$ $M_{\sun}$, the increase
 in momentum due to star formation is roughly estimated to be 
$\sim 13$ $M_{\sun}$ km s$^{-1}$.
The estimated increase in momentum is much smaller than the outflow momentum 
supplied by massive protostar(s).
For example, the total outflow momentum supplied by massive protostar(s) 
with a luminosity of $\sim 10^4$ $L_{\sun}$ is estimated to be 
$\sim 100$ $M_{\sun}$ km s$^{-1}$ (Zhang et al. 2005).
This indicates that at most only a tenth the total outflow momentum is injected 
to the present core, and almost all of the outflow momentum is likely to be 
supplied into the intercore gas or outside the clump.
In other words, massive-star-forming cores should have a large line width
even prior to star formation. Therefore, we suggest that although star formation
activity tends to enhance the line width of the cores, massive starless cores 
should form from a dense gas with a large kinetic motion.

Finally, considering that the range of the radius of the cores is small,
the relationship between the mass and the line width of the cores suggests 
that the average H$_2$ density of the cores would increase with the line width 
of the cores.
According to this result and the discussion on the external pressure in 
$\S$ 4.1.2,
a small dissipation of the turbulent motion is sufficient for a core formed 
in the center of the clump with the H$_2$ density structure to be 
gravitationally bound because the initial condition of this core necessarily 
has a high H$_2$ density and a high external pressure. 
In this case, the index $\alpha$ of the line width--radius relationship from 
the clump to the core would become very small ($\sim 0.06$).
However, a core formed at the edge of the clump must undergo a large dissipation
of turbulent motion to be gravitationally bound, and the index $\alpha$ would 
become large ($\sim 0.9$).
In other words, the physical parameters of the core would depend on the place 
where the core is formed in the clump.
This relationship is discussed in detail in $\S$ 4.2.2.

\subsection{Relationship among Young Stars, Dense Cores, and Dense 
Clumps}
Here, we discuss the relationships between the mass of the star formed in 
the core and the physical parameters of the core and between the physical 
parameters and the positions of the cores in the clump. 
Generally, the mass of a star is expected to depend on the physical properties 
of the parental core in a clump. Thus, the characteristics of the stellar 
member of the cluster would depend on the characteristics of the cores in 
the clump. In addition, Saito et al. (2007) found that the maximum mass of 
the stars associated with the clump increases with the line width of the clump 
and that a good correlation exists between the stellar number density of 
the cluster and the average H$_2$ density of the clump.
These results indicate the importance of revealing the relationships among 
the clump, the cores, and young (proto)star(s) to investigate both cluster 
formation and massive star formation.

\subsubsection{Relationship between the Mass of Formed Stars and the Physical 
Properties of Cores}
Typical massive-star-forming cores have a higher H$_2$ density, larger line
width, and higher temperature than the low-mass star-forming cores
(e.g., Churchwell et al. 1992; Olmi et al. 1993). 
Although many observations with such tendencies have been reported, 
the relationship between the mass of the star and the physical properties of 
the parental core has never been quantitatively investigated. 
The main reason is that the discovery of the cores forming stars with 
various masses in the same region has been very difficult. 
However, typical stellar clusters have many stars with various masses, and 
stars formed in star-forming cores described here are expected to have 
a wide range of masses.
Therefore, we estimated the stellar masses of the stars using the same method 
as Saito et al. (2007). For stars whose luminosities have been estimated or 
stars with centimeter-continuum sources, we estimated their spectral types 
using these parameters. We estimated the spectral types of other stars from 
the $H-K_{\rm S}$/$K_{\rm S}$ color--magnitude diagrams using the 2MASS catalog.

Figure 11 shows the relationships between the mass of (proto)stars associated 
with cores, $M_{\ast}$, and the physical parameters, LTE mass and line width, 
of the cores. 
A tendency exists for the mass of stars to increase with these parameters. 
These results indicate that the mass of the star would depend on the mass 
(or the H$_2$ density) and internal kinetic motion of the parental core. 
Generally, high average H$_2$ density and large internal kinetic motion 
would be expected to lead to a high mass accretion rate 
(e.g., McKee \& Tan 2003).
Although the plot of the relationship between the mass of the stars and the mass
of the cores has a large scattering, the plot can roughly be expressed by 
the relationship of $M_{\ast}$ = $M_{\rm core}$, which is indicated by the solid 
line. From this relationship, the star-formation efficiency (SFE) is roughly 
constant at $\sim 50$\% using the equation of 
SFE = $M_{\ast}$/($M_{\ast}$ + $M_{\rm core}$).
Note that the mass of the core is expected to decrease with the age of 
the protostar(s).

Based on the above relationships and the mass--line width relationship 
in $\S$ 4.1.3, the magnitude of internal kinetic energy of the core would 
determine the mass of the formed stars. 
In addition, the core must have existed in the clump at least during 
the lifetime of the clump for star formation to occur therein. 
It is therefore necessary for the core to be gravitationally bound.
From these results, we suggest that the necessary and sufficient condition
to form a higher mass star is that the core has a larger internal kinetic 
motion and is gravitationally bound. In this case, it is not necessary for 
the core to be self-gravitationally bound if high external pressure is brought 
to bear on the core surface. Because the internal kinetic motion basically 
decreases with time, a structure such as a molecular cloud much larger than 
a core or a clump must previously have had a large internal kinetic motion
and must be gravitationally bound for massive star formation to occur. 
Thus, GMCs and clouds with an external effect, for example, 
H\,{\footnotesize II} regions and supernova remnants, would be good 
environments for massive star formation because such clouds would have 
a large external pressure and be supplied with large kinetic energy.

\subsubsection{Relationship  between the Cores and the Clumps}
Here, we discuss the spatial distribution of the physical parameters of 
the cores in the clump. In $\S$ 4.1, we found that dense gas with a large 
kinetic motion in a clump seems to be necessary to form massive cores, 
and the external pressure on the surface of cores must be large for 
massive cores to achieve an equilibrium state. In addition, cores with 
large external pressure are concentrated in the center of the clump.
Thus, the physical properties of the core are expected to depend on 
the physical condition of the region where the core is formed in the clump.

First, we examined the relationship between the physical parameters,
line width and LTE mass, of the cores and the core distance (see $\S$ 4.1.2). 
Figure 12(a) shows the masses of the cores plotted against the core distance. 
Although no clear tendency is seen in this plot, the upper limit of 
the core mass would gradually decrease with the core distance. 
If the average H$_2$ density of the cores is proportional to the H$_2$ density 
in the clump having a density structure of $\sim \,R_{\rm cen}\,^{-\gamma}$, 
at the region where the core was formed, the core mass can be expressed 
by the relationship 
\begin{equation}
M_{\rm core} = \frac{4\,\pi}{3}\,R_{\rm core}\,^{3}\,\mu'\,m_{\rm H}\,
n({\rm H_{2}})_{\rm core} = \frac{4\,\pi}{3}\,R_{\rm core}\,^{3}\,\mu'\,m_{\rm H}\,
C\,n_0\,(\frac{R_{\rm cen}}{R_{\rm CL}})^{-\gamma}, 
\end{equation}
where $\mu'$ and $m_{\rm H}$ are the mean molecular weight per one H$_2$ 
molecule and the hydrogen mass, respectively, and $n_0$ and C are the H$_2$ 
density in the clump at the distance of $R_{\rm CL}$ from the center of 
the clump and the proportional coefficient, respectively.
Generally, the H$_2$ density structure of a clump is suggested to have 
a relationship with ($R_{\rm cen}$/$R_{\rm CL})^{-\gamma}$ 
($\gamma \sim 1.0$--2.0; e.g., van der Tak et al. 2000).
Here, if a clump has a mass of 300 $M_{\sun}$ and a radius of 0.3 pc and 
$R_{\rm core}$ and C are 0.03 pc and 50, which are typical respective values 
for the present sample, the mass--distance relationships with 
$\gamma$ = 1.0 and 2.0 are indicated by the solid and dashed lines in 
Figure 12(a), respectively.
As shown in Figure 12(a), the distribution of the upper limit of the core mass 
is roughly similar to the relationships expressed by the two lines. 
From this result, the mass (or the average H$_2$ density) of the core is 
suggested to depend on the H$_2$ density of the region where the core is 
formed in the clump, and the distribution of the core mass is suggested to be 
controlled by the H$_2$ density structure in the clump.

Figure 12(b) shows the ratio of the line width for the core and the clump 
plotted against the core distance. The starless cores with a high line width 
ratio ($> 0.7$) are concentrated in the center of the clump, and the upper 
limit of this ratio would gradually decrease with the core distance. 
The reason for this feature would be the change in both the core mass and 
the external pressure on the core surface.

Here, if we assume the cores are gravitationally bound by the clump pressure, 
we can estimate the maximum line width of the cores using the virial equation 
(see $\S$ 4.1.2).
If the clump with the physical parameters adopted in the above discussion is 
gravitationally bound, a line width of the clump is 2.2 km s$^{-1}$. 
In this case, the ratio of the maximum line width of the cores to the line width
of the clump with $\gamma$ = 1.0 and 2.0 is indicated by the solid and dashed 
lines, respectively, in Figure 12(b).
As shown in Figure 12(b), the distribution of the upper limit of the line width
ratio is similar to the relationships expressed by the two lines.
As we have seen in $\S$ 4.1.2 and the above-mentioned mass--distance 
relationship, the external pressure on the core surface and the core mass are 
likely to depend on the core distance due to the H$_2$ density structure in 
the clump.
Therefore, to bind gravitationally, the internal kinetic motion of the core 
would also necessarily depend on the core distance.

From the discussions in this subsection, we suggest that the central region 
of the clump, which has a density structure of $\sim \,R_{\rm cen}\,^{-\gamma}$, 
is the environment in which the core with a large internal kinetic motion and 
a high average H$_2$ density is easily formed and that the internal kinetic 
motion and the average H$_2$ density of the core would depend on 
the H$_2$ density structure in the clump.
In addition, according to the result in $\S$ 4.2.1, the mass of a star formed 
in the core would decrease with the distance from the center of the clump. 
Therefore, the physical characteristics of the cluster would depend on 
the H$_2$ density structure, internal kinetic motion, and the mechanism of 
core formation in the clump.

\subsection{Mechanism of Cluster Formation}
Finally, we discuss the relationship between cluster formation and the dense gas
structures, cores and clumps. As we have seen in $\S$ 4.1 and $\S$ 4.2, 
many cores with a size of $\sim 0.03$ pc that are formed in a clump with 
a size of $\sim 0.3$ pc have a wide range of the line width and the mass, 
and the spatial distribution of the physical parameters of the cores would 
depend on the H$_2$ density structure in the clump. 
In addition, the mass of the star would depend mainly on the internal kinetic 
motion of the parental core.
The stellar number density of the cluster is mentioned as an important 
characteristic of a cluster. The stellar number density of the cluster is 
generally much higher than that of typical low-mass star-forming regions.
In this regard, Saito et al.(2007) found a good relationship between the number 
density of YSOs associated with the clump and the average H$_2$ density of 
the clump, and they suggested that it is necessary for a clump to have 
a high average H$_2$ density for cluster formation to occur therein.
This suggests that a relationship may exist between the number density
of the cores in the clump and the average H$_2$ density of the clump.
First, we estimated the number density of the cores to investigate 
this relationship.

Figure 13(a) shows the relationship between the number density of the cores in 
the clump, $n_{\rm core}$, and the average H$_2$ density of the clump, 
$\overline{{\rm n(H_2)}_{\rm clump}}$.
A good correlation exists between the number density of the cores and 
the average H$_2$ density of the clump. 
This relationship is roughly expressed by the equation $n_{\rm core} \sim$ 
$\overline{{\rm n(H_2)}_{\rm clump}}$\,$^{1.9}$ and the index of this relationship
is similar to the index ($\sim 2.0$) of the relationship between the number 
density of the YSOs and the average H$_2$ density of the clump obtained by 
Saito et al. (2007).
This indicates that star formation in the clump is based on core formation in 
the clump.
Although these relations are not expressed by the mechanism of core formation
using simple gravitational instability as proposed by Saito et al. (2007), 
we suggest that for core formation with a high number density of the cores,
it is necessary for the clump to have a high average H$_2$ density as 
an initial condition.

Next, we examine the relationship between the number of the cores and 
the YSOs to reveal the timescale of the cores.
Figure 13(b) shows the number of YSOs, $N_{\ast}$, associated with the clump, 
which were identified by Saito et al. (2007), plotted against the number of 
cores in the clump, $N_{\rm core}$. 
The numbers of the YSOs are one to three times larger than those of the cores.
Considering that almost all of the YSOs would be low-mass stars, the lifetime
of the YSO would be $\sim 1 \times 10^7$ yr (e.g., Lada 1999).
In this case, the lifetime of the core is roughly estimated to be 
$\sim 3 \times 10^{6}$ yr if only one star is formed per core. 
In addition, we find 27 star-forming cores in this study and 10 in 
Saito et al. (2006), 
and the number corresponds to $\sim 19$\% of all cores.
This result indicates that the lifetime of the star-forming cores, which can 
be detected by our interferometric observation, is $\sim 6 \times 10^5$ yr.
This timescale is consistent with the fact that the typical mass of the cores 
associated with class II stars, whose lifetime is $\sim 10^{6}$ yr 
(e.g., Lada 1999), was roughly estimated to be $\ll 1 M_{\sun}$ in low-mass 
star forming regions (e.g., Tsukagoshi et al. 2007) because the detection 
limit of the mass of the cores is $\sim 1 M_{\sun}$.
Therefore, most of the YSOs in the clump are not associated with the cores 
with a mass of $> 1 M_{\sun}$ and the YSOs associated with these cores are 
regarded as young protostars.

Finally, we look at the core-formation efficiency (CFE) of the clump.
Figure 13(c) shows CFEs ($\Sigma M_{\rm core}$/$M_{\rm clump}$, where 
$M_{\rm core}$ and $M_{\rm clump}$ are the mass of the core in the clump and 
the mass of the clump, respectively) plotted against the number density of 
cores in the clump. Although no strong correlation is seen between the CFEs and 
the number density of cores, the CFE gradually increases with the number 
density of cores in the clump. 
Considering the relationship between the number density of cores and 
the average H$_2$ density of the clump in Figure 13(a), this feature indicates 
that the dense gas in a clump with a high average H$_2$ density is efficiently 
converted into many cores.
From the results that the CFE is $\sim 3$--60\% and that the lifetime of 
the core is $\sim 3 \times 10^{6}$ yr in the above discussion, 
the conversion rate from the dense gas to cores per $\sim 1 \times 10^6$ yr 
(core-formation rate : CFR) in the clump is estimated to be 
$\sim 1$ -- 20\%/$10^6$ yr.
Thus, in a clump with a low average H$_2$ density, core formation with 
a low CFR ($\sim$ a few \%/$10^6$ yr) could have occurred over a long time of 
$> 10^7$ yr if the CFR is always constant. However, only a part of the dense 
gas in the clump would be converted to cores because this timescale is longer 
than the typical lifetime of the GMC 
($\sim 10^{7}$ yr; e.g., Blitz et al. 2007).
On the contrary, in a clump with a high average H$_2$ density, 
core formation with a high CFR ($\ge 10$\%/$10^6$ yr) could have occurred 
over a short time of $\le 10^7$ yr. This indicates that most of the dense gas 
in the clump would be converted to cores because this timescale is shorter than 
the typical lifetime of the GMC.

From the above considerations and the discussions in $\S$ 4.1 and 4.2,
we determined that the mechanism of cluster formation would strongly depend on
the mechanism of core formation in a clump and that the mechanism of core
formation would be controlled by both the H$_2$ density structure and 
the kinetic motion in the clump. Moreover, the characteristics of the cluster 
would be determined by the distribution of the physical parameters of cores in 
the clump.
If a clump with a very high average H$_2$ density ($\sim 10^6$ cm$^{-3}$), 
which has a size scale of $\sim 0.3$ pc, is formed in a GMC, this clump 
would be expected to form a cluster with a very high stellar number density 
very efficiently in a short time based on the discussion of the CFR.
In addition, considering that the virial mass of the clump is similar to 
the LTE mass (Saito et al. 2007), the line width of the clump is estimated to 
be $\sim 10$ km s$^{-1}$ using the average H$_2$ density and the radius.
According to the line width--radius relationship in $\S$ 4.1.1 and 
the stellar mass--line width relationship in $\S$ 4.2.1, the cluster formed in 
the clump would include very massive (e.g., O3) stars. 
Thus, the clumps with a very high average H$_2$ density ($\sim 10^6$ cm$^{-3}$)
would be expected to form a cluster with both a very stellar number density and
very massive stars. In addition, if many clumps are formed in a GMC, 
star-forming regions within the GMC will include many dense clusters with 
very massive stars and will be extended with a size of $\sim 100$ pc because 
the typical size of the GMC is $\sim 100$ pc.
Indeed, very active star-forming regions occurring in external galaxies, 
for example, 30 Doradus region in LMC, are extended with a size scale of 
$\ge 100$ pc and have many dense clusters, which contain many O3 stars, with 
a size scale of $\sim 0.3$ pc, for example, R136a core in 30 Doradus.
Therefore, we suggest that the mechanism of cluster formation in such active
star-forming regions would be basically similar to that in our galaxy, although
the clumps in such active star-forming regions would have much higher H$_2$
density and much larger kinetic motion than those in our galaxy.

\section{CONCLUSION}
We made high-angular resolution observations with the Nobeyama Millimeter 
Array (NMA) of six cluster-forming regions including massive (proto)stars, 
in the $J\!=\!1$--$0$ transition of C$^{18}$O molecular emission line and 
98 GHz and 110 GHz continuum emission.
We identified 171 C$^{18}$O cores in 12 clumps toward six cluster-forming 
regions including massive (proto)stars. Based on our observations, we studied 
the correlation between several physical quantities of the C$^{18}$O cores 
and clumps, and discussed the formation mechanism of star clusters.
The main results of the present study are summarized as follows:

1. The ranges of the radius, line width, and molecular mass of the cores are 
   0.01--0.09 pc, 0.43--3.33 km s$^{-1}$, and 0.5--54.1 $M_{\sun}$, 
   respectively.

2. We identified 13 millimeter continuum sources (MCSs). Seven are associated 
   with 2MASS sources and/or centimeter continuum sources (CCSs). Together with 
   the one MCS that shows similar characteristics to the above seven MCSs, 
   these eight MCSs would include massive protostars. 
   The range of the mass of MCSs is 2.7--111 $M_{\sun}$. 
   In addition, we compared C$^{18}$O cores with MCSs, 2MASS sources, and CCSs. 
   We identified 27 star-forming cores including eight massive-star-forming 
   cores.

3. Some cores have various line widths in a single clump. The range of 
   the index of the relationship between the line width and radius of the core 
   and the parental clump differs from core to core. This implies that 
   the degree of dissipation of the turbulent motion varies within a single 
   clump.

4. For most cores, the virial masses are about twice larger than the LTE masses.
   This indicates that external pressure is necessary for the cores to be 
   gravitationally bound.
   Based on the virial analysis, the distribution of the external pressure
   required to bind the cores is likely to be comparable to that of 
   the pressure in the clump with an H$_2$ density structure of 
   $\sim r^{-\gamma}$ ($\gamma \sim 1$--2).

5. The mass of the cores increases with the line width. This indicates that 
   the average H$_2$ density of the cores would depend on the kinetic motion 
   of the dense gas in the core because the range of the radius of the core 
   is small.

6. The mass of the star depends on the line width and the mass of the parental 
   core. In addition, we found that the star-formation efficiency (SFE) is 
   roughly constant at $\sim 50$\% and that no clear dependence exists between 
   the SFE and the mass of the formed stars.

7. The mass, line width, and external pressure of the core decrease with 
   distance from the center of the clump, and a relationship exists between 
   these decreases and the H$_2$ density structure of the clump.
   We suggest that the spatial distribution of physical parameters of the cores
   strongly depends on the H$_2$ density structure in the clump.

8. The number density of the cores and the number density of the young (proto) 
   stars in the clump have a similar relationship to the average H$_2$ density 
   of the clump. In addition, the core-formation rate (CFR) of the clump is 
   estimated to be $\sim 1$--20\% per $\sim 1 \times 10^{6}$ yr and the CFR 
   gradually increases with the average H$_2$ density of the clump. 

\acknowledgments

We thank to the staff of the Nobeyama Radio Observatory 
(NRO) for both operating the Millimeter Array and helping us with data 
reduction.
We thank to Kazuyoshi Sunada, Tomofumi Umemoto,  and Takeshi Sakai and also
to the referee for their valuable comments. 
This publication makes use of data produced by 2MASS, which is a joint 
project of the University of Massachusetts and Infrared Processing and 
Analysis Center funded by NASA and NSF. This study also made use of 
the SIMBAD database.





\clearpage



\onecolumn
\begin{figure}
\begin{center}
\epsscale{0.6}
\plotone{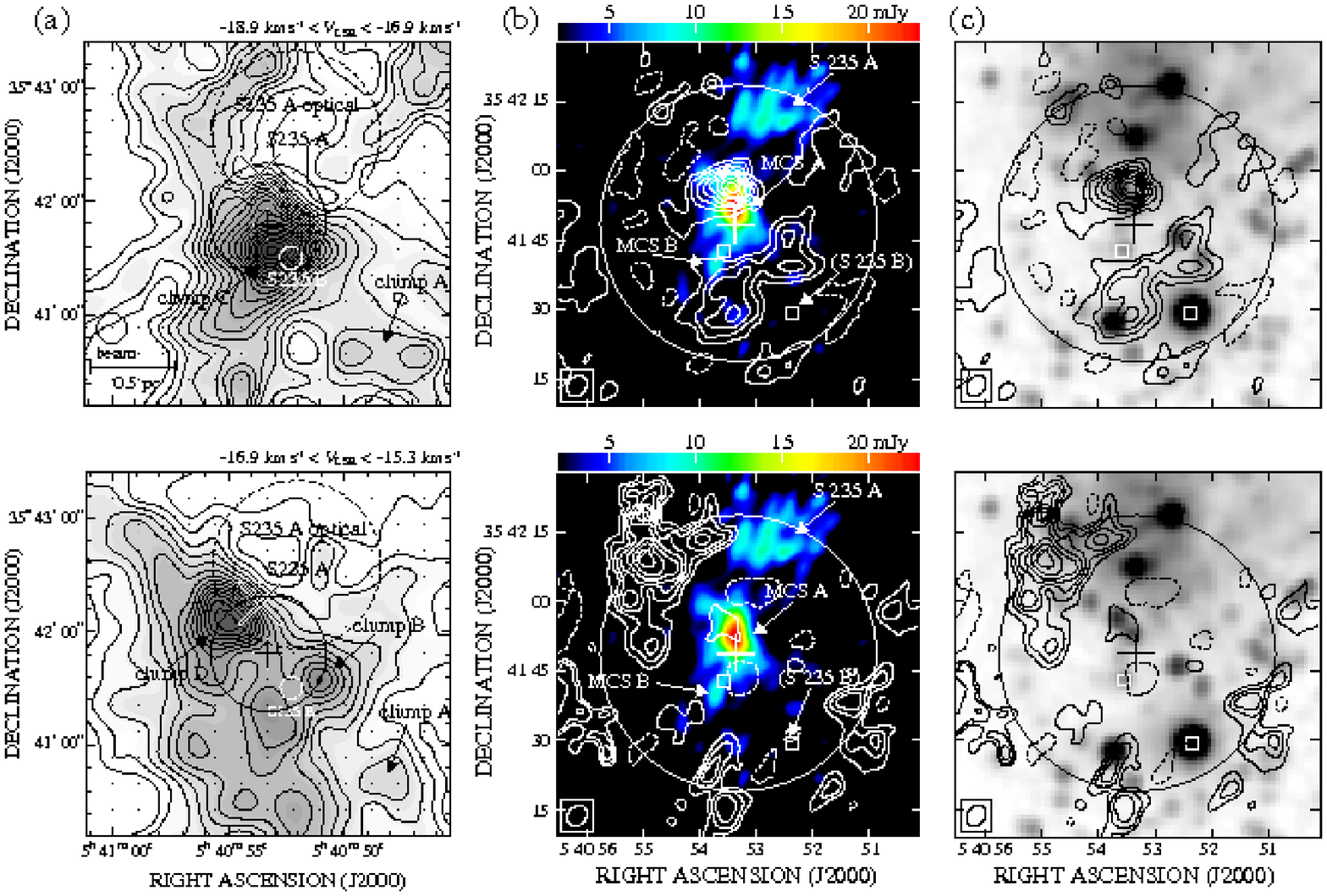}
\epsscale{0.50}
\plotone{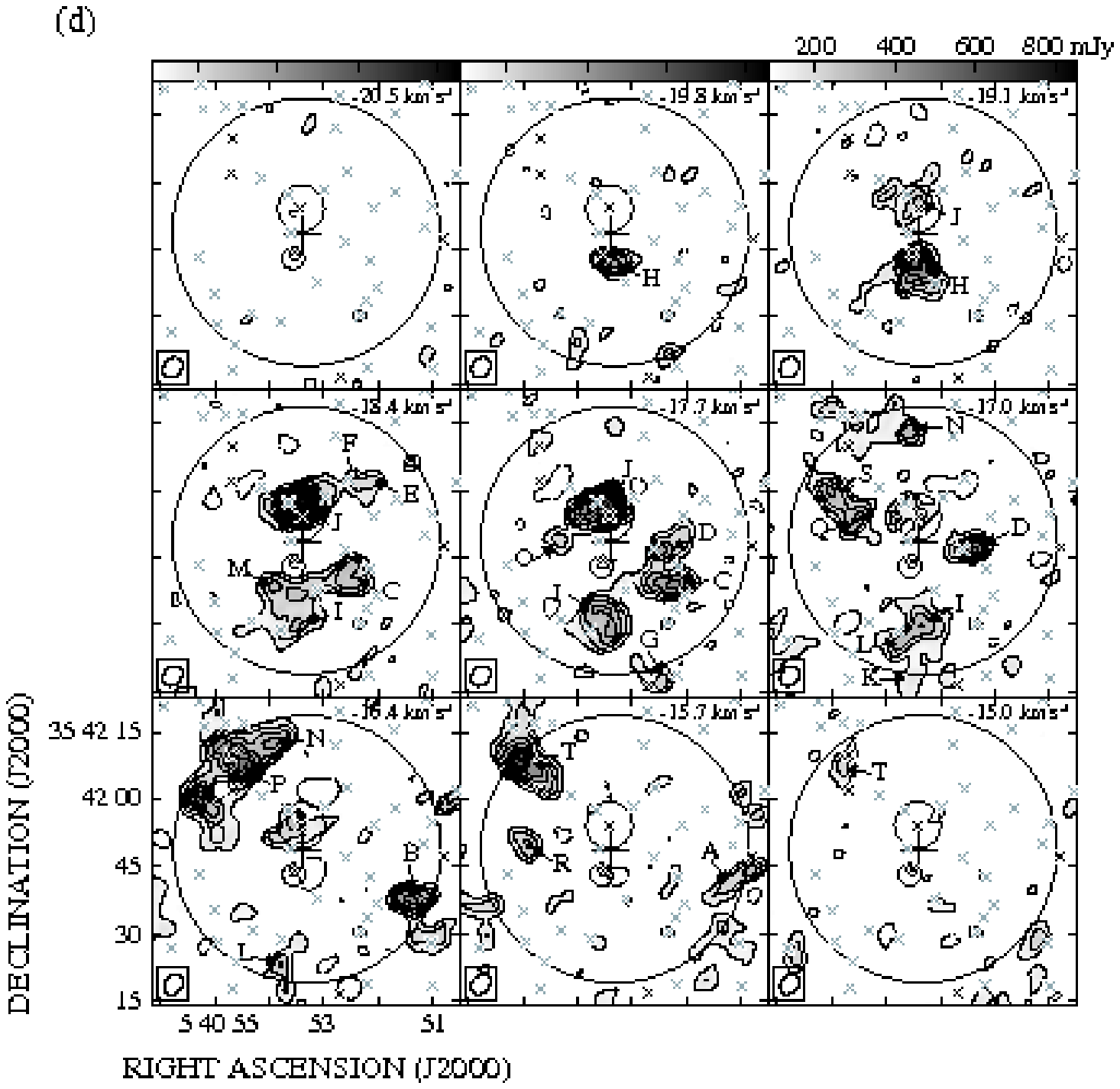}
\caption{
(a) The distribution of the C$^{18}$O (J$\,$=$\,$1--0) emission toward
    IRAS\,05375+3540 and the clumps obtained by Saito et al. (2007) are shown.
    The integrated velocity ranges for the upper and lower panels are between 
    $-18.9$ and $-16.9$ km s$^{-1}$ and between $-16.9$ and 
    $-15.3$ km s$^{-1}$, respectively. 
    The large solid circle, the plus sign, and the squares denote the primary 
    beam of NMA, the position of the maser source, and the CCS, respectively.
    The large cross and the dashed circles indicate the IRAS source 
    and the H\,{\footnotesize II} region, respectively.
(b) Color images of the 110 GHz continuum emission overlaid with contours of 
    C$^{18}$O (J$\,$=$\,$1--0) emission obtained by the present observation. 
    The integrated velocity is the same as in Figure 1(a). 
    Contour levels are from $-6 \sigma$ in steps of 3 $\sigma$, except for 
    0 $\sigma$. The 1 $\sigma$ noise levels are 24 mJy for both panels. 
(c) Gray images of the 2MASS $K_{\rm s}$ band overlaid with contours of 
    C$^{18}$O emission.
(d) Channel maps of C$^{18}$O emission and the cores obtained by the present 
    observation are shown. The LSR velocity is marked at the top right corner
    in each panel. 
    The small circles and the crosses indicate the MCSs and 2MASS sources, 
    respectively. In particular, the black crosses indicate the 2MASS source 
    associated with the C$^{18}$O core.
    Contour levels are from $-6 \sigma$ in steps of 2 $\sigma$ and 
    the 1 $\sigma$ noise levels are 34 mJy.}
\label{fig.1}
\end{center}
\end{figure}

\clearpage

\begin{figure}
\begin{center}
\epsscale{0.70}
\plotone{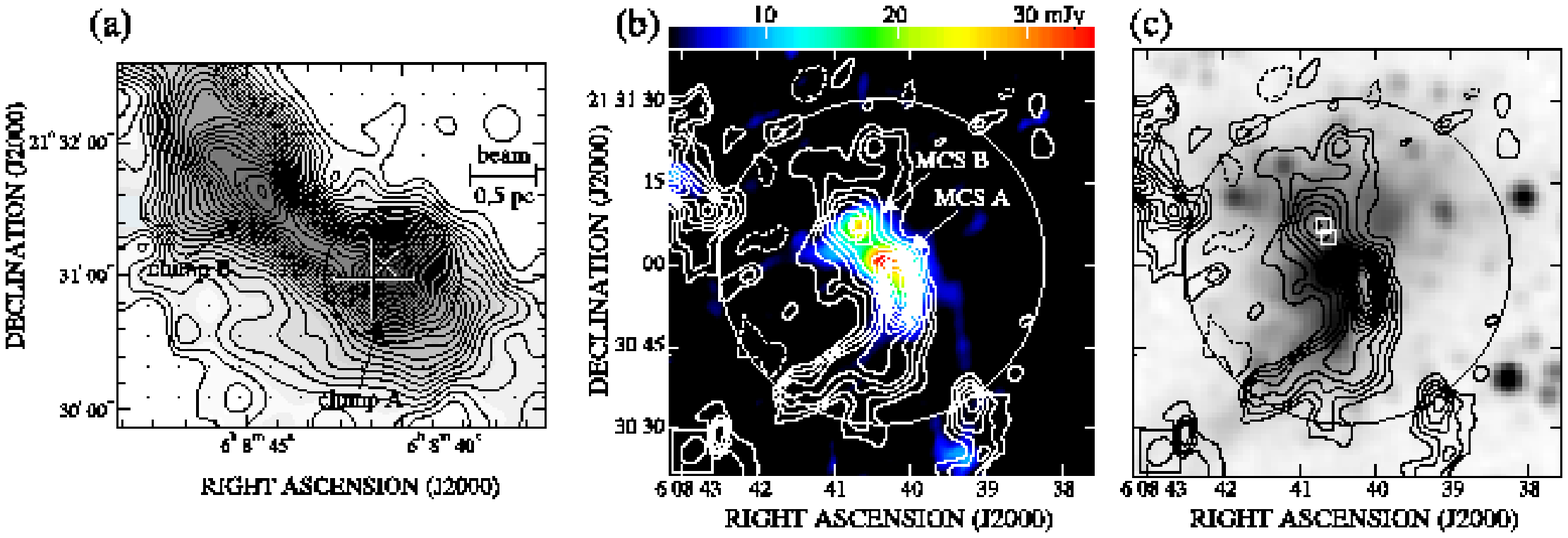}
\epsscale{0.50}
\plotone{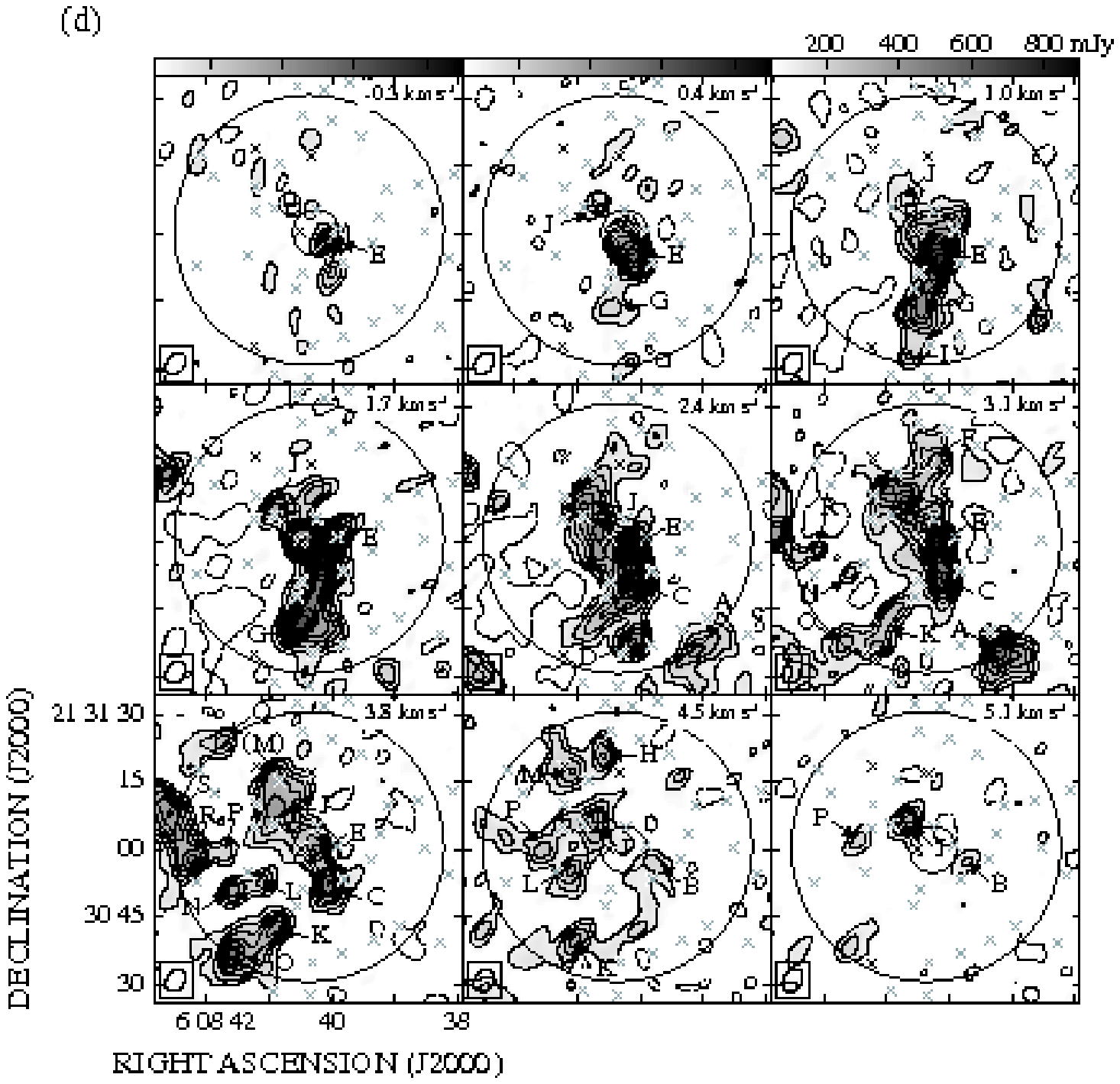}
\caption{
(a) Same as in Figure 1(a), but for IRAS\,06056+2131. The integrated velocity 
    range is between $-1.5$ and $6.5$ km s$^{-1}$.
(b) Same as in Figure 1(b), but the 1 $\sigma$ noise level is 12 mJy. 
(c) Same as in Figure 1(c). 
(d) Same as in Figure 1(d), but the 1 $\sigma$ noise levels are 30 mJy.}
\label{fig.2}
\end{center}
\end{figure}

\clearpage 

\begin{figure}
\begin{center}
\epsscale{0.6}
\plotone{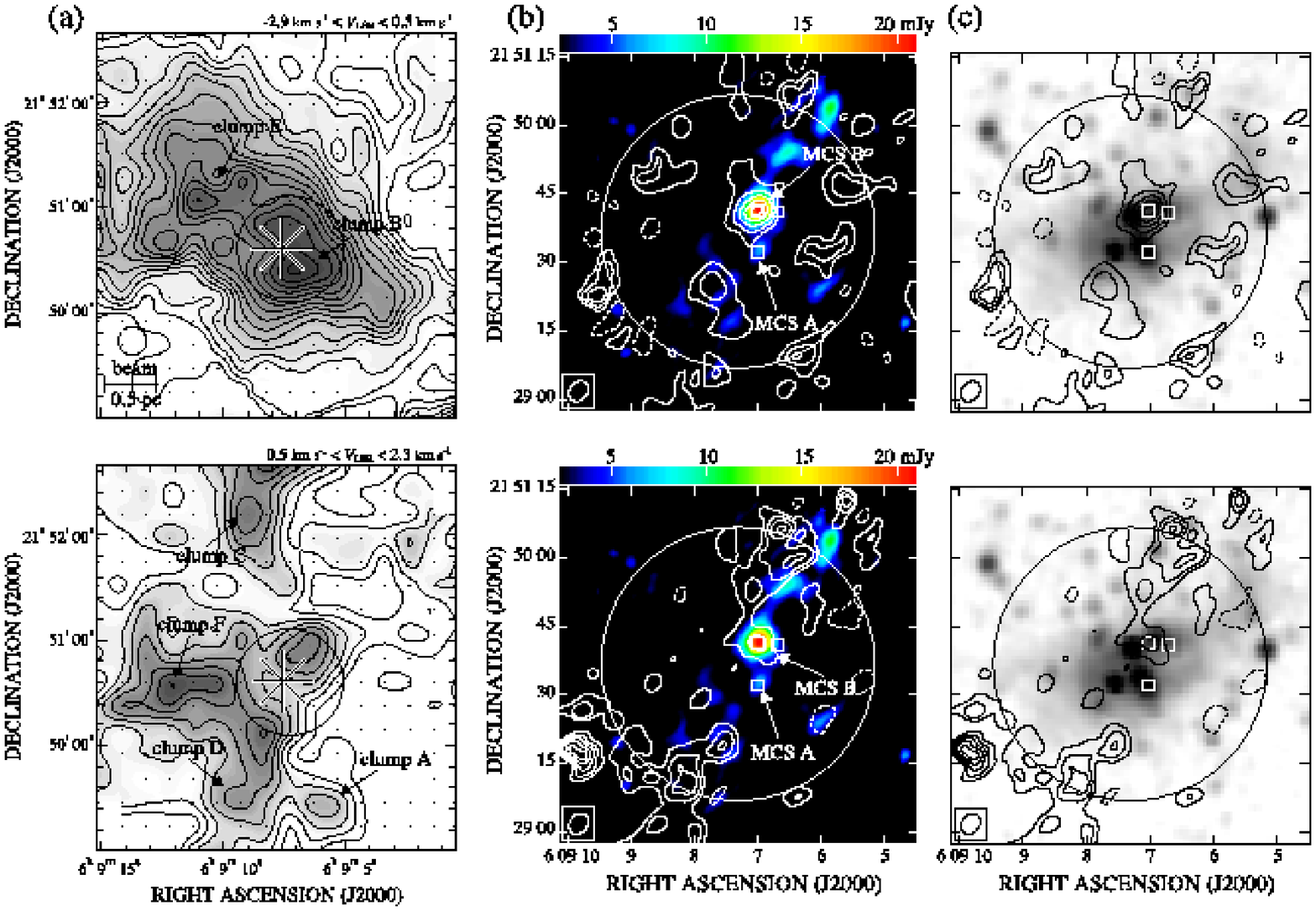}
\epsscale{0.5}
\plotone{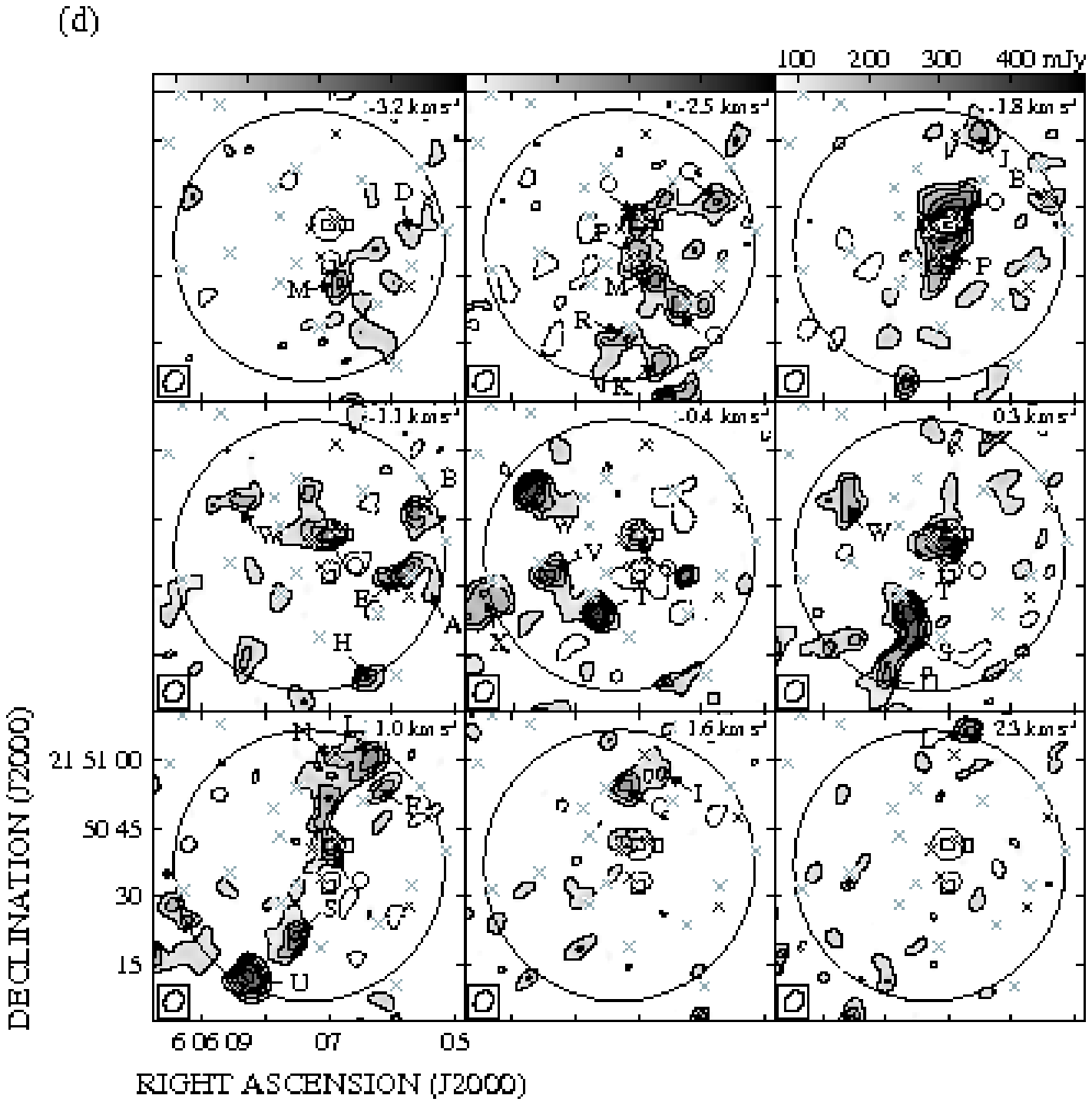}
\caption{
(a) Same as in Figure 1(a), but for IRAS\,06061+2151. The integrated velocity 
    ranges for the upper and lower panels are between $-2.9$ and $0.5$ 
    km s$^{-1}$ and between $0.5$ and $2.3$ km s$^{-1}$, respectively.
(b) Same as in Figure 1(b), but the 1 $\sigma$ noise levels are 18 mJy 
    (upper panel) and 21 mJy (lower panel). 
(c) Same as in Figure 1(c). 
(d) Same as in Figure 1(d), but the 1 $\sigma$ noise levels are 30 mJy.}
\label{fig.3}
\end{center}
\end{figure}

\clearpage 

\begin{figure}
\begin{center}
\epsscale{0.6}
\plotone{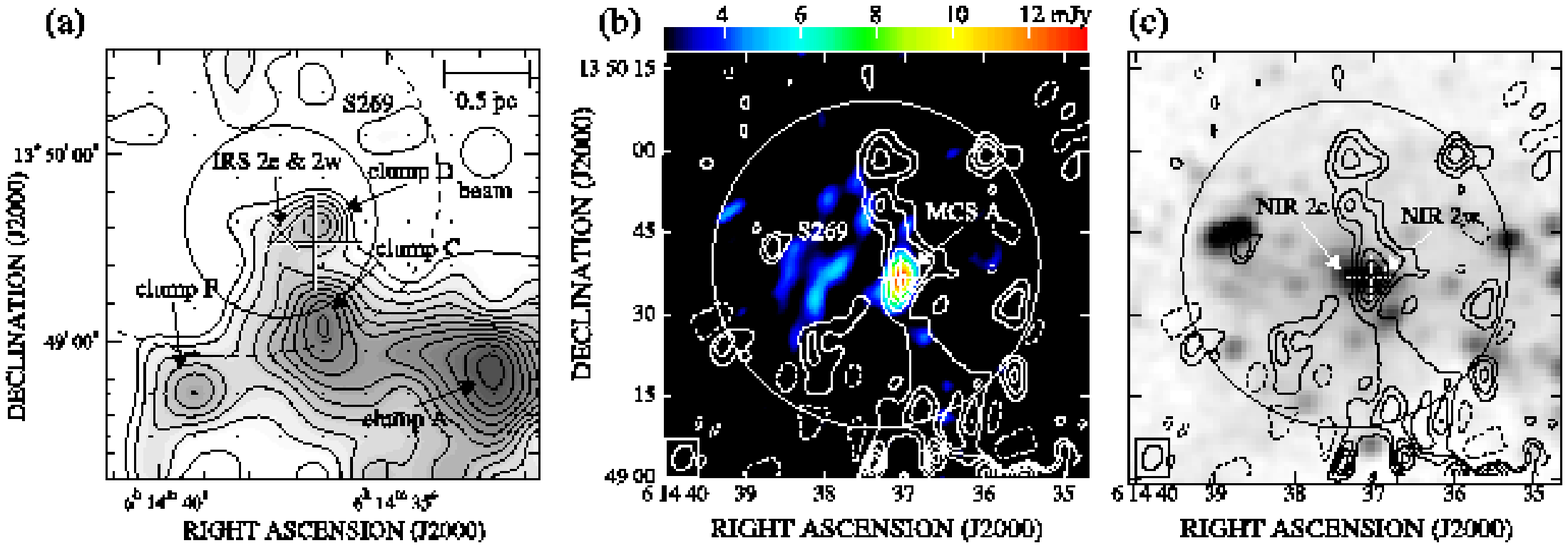}
\epsscale{0.50}
\plotone{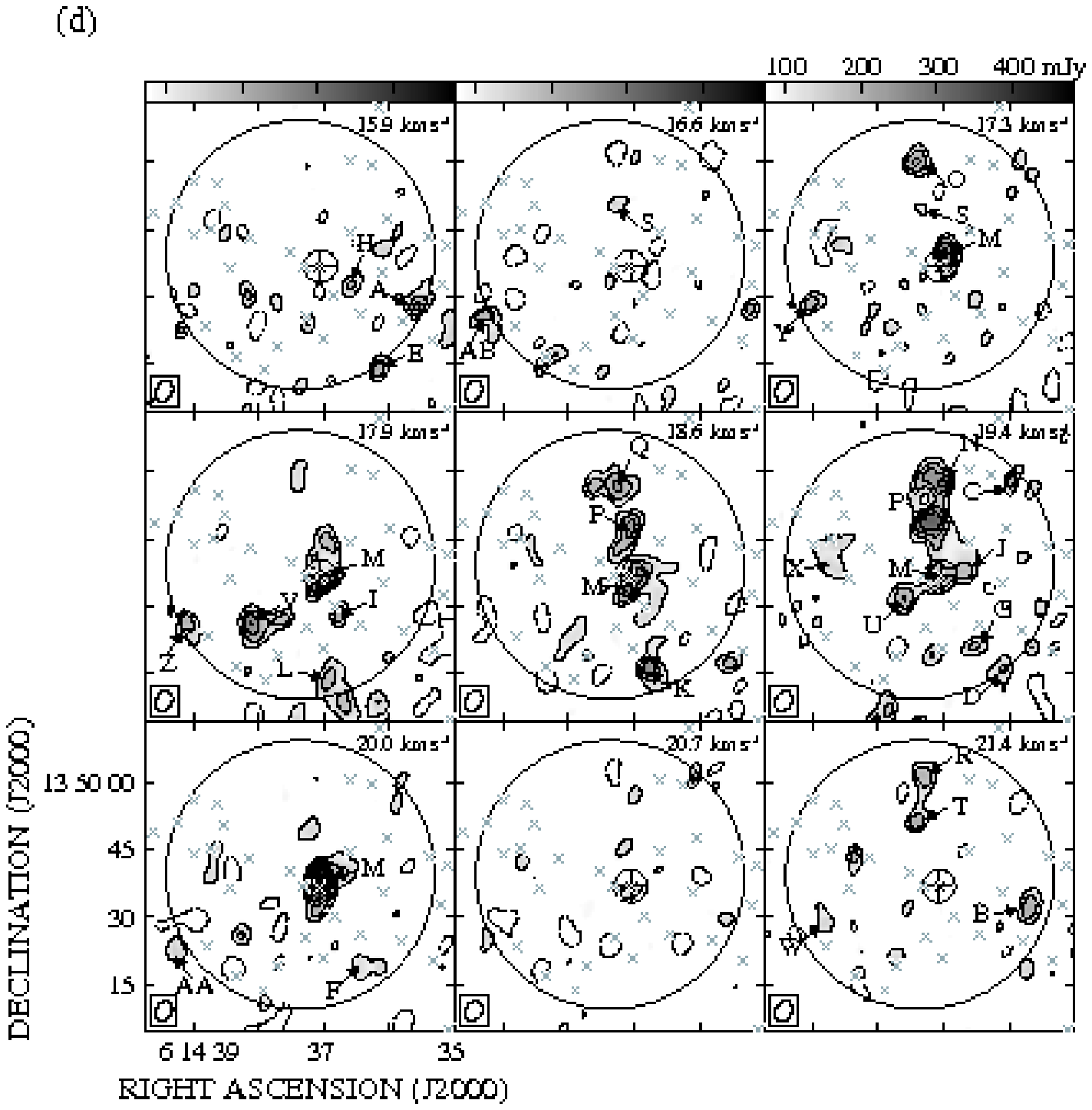}
\caption{
(a) Same as in Figure 1(a), but for IRAS\,06117+1350. The integrated velocity 
    range is between $14.5$ and $23.0$ km s$^{-1}$. 
    The cross indicates NIR 2e \& 2w.
(b) Same as in Figure 1(b), but the color image is of the 98 GHz continuum 
    emission and the 1 $\sigma$ noise level is 10 mJy. 
(c) Same as in Figure 1(c). 
(d) Same as in Figure 1(d), but the 1 $\sigma$ noise levels are 30 mJy.}
\label{fig.4}
\end{center}
\end{figure}

\clearpage 

\begin{figure}
\begin{center}
\epsscale{0.6}
\plotone{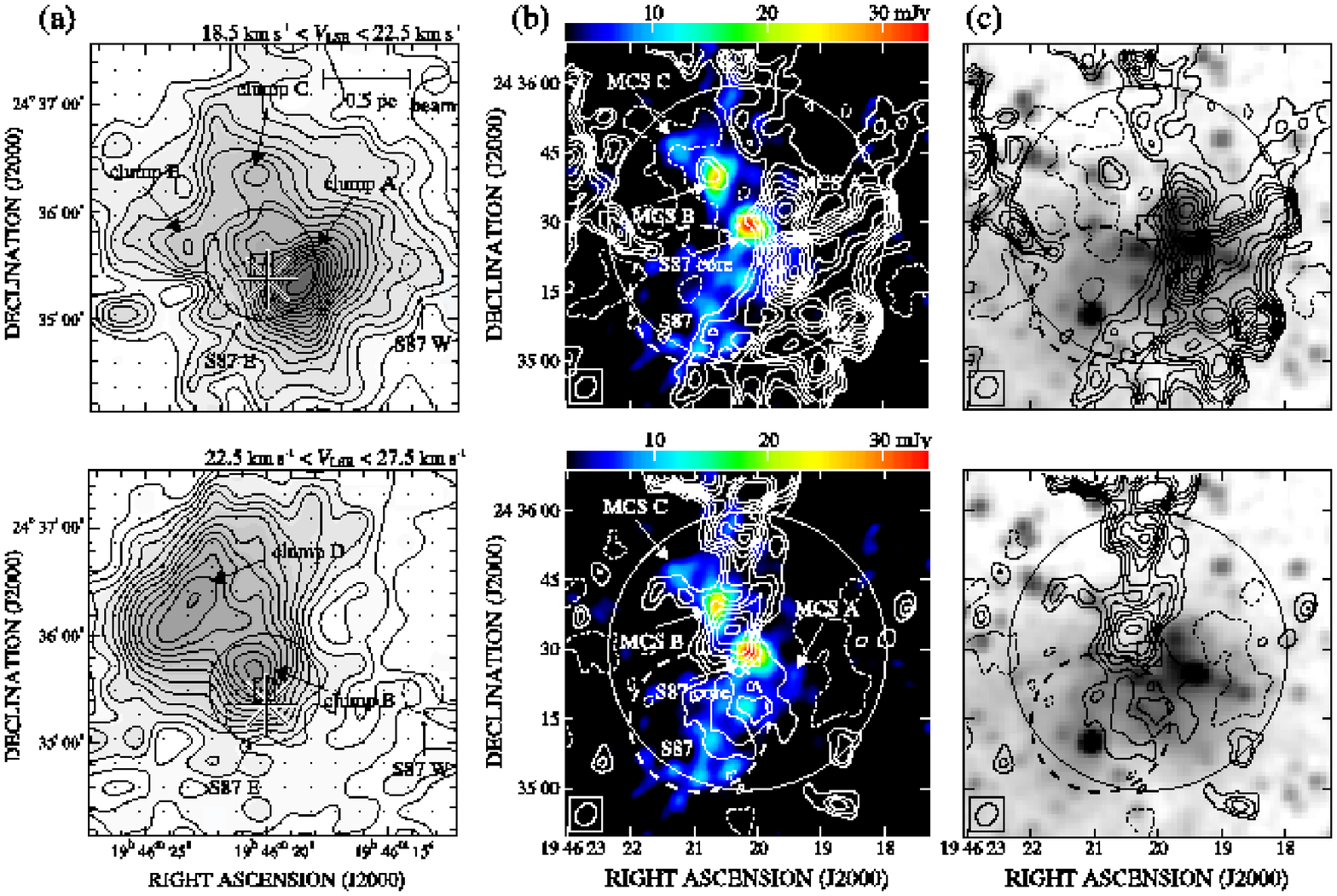}
\epsscale{0.5}
\plotone{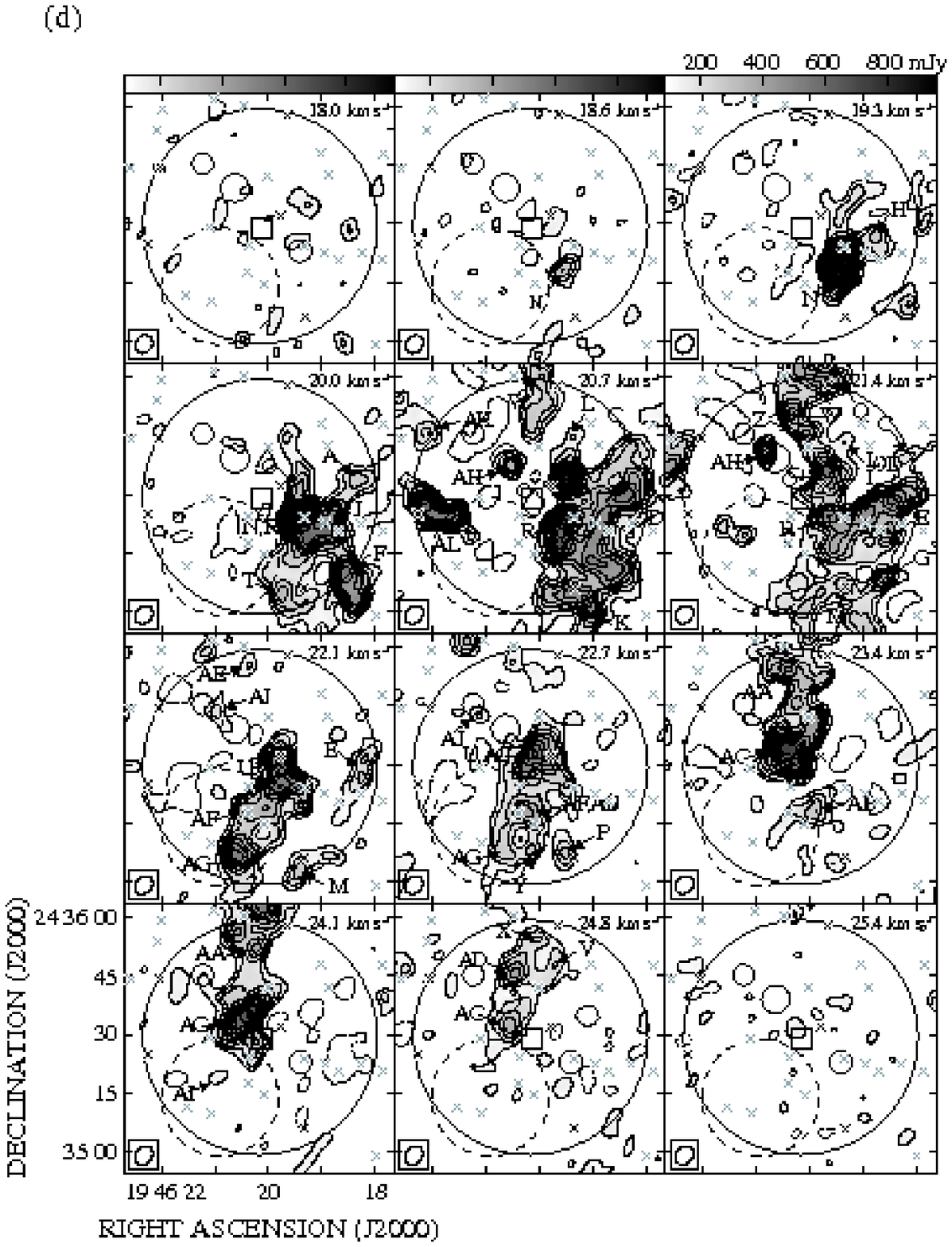}
\caption{
(a) Same as in Figure 1(a), but for IRAS\,19442+2427. The dashed circles 
    denote the area of the radio H\,{\footnotesize II} regions. The integrated 
    velocity ranges for the upper and lower panels are between $18.5$ and 
    $22.5$ km s$^{-1}$ and between $22.5$ and $27.5$ km s$^{-1}$, respectively.
(b) Same as in Figure 1(b), but the 1 $\sigma$ noise levels are 14 mJy 
    (upper panel) and 12 mJy (lower panel).
(c) Same as in Figure 1(c). 
(d) Same as in Figure 1(d), but the 1 $\sigma$ noise levels are 30 mJy.}
\label{fig.5}
\end{center}
\end{figure}

\clearpage 

\begin{figure}
\begin{center}
\epsscale{0.6}
\plotone{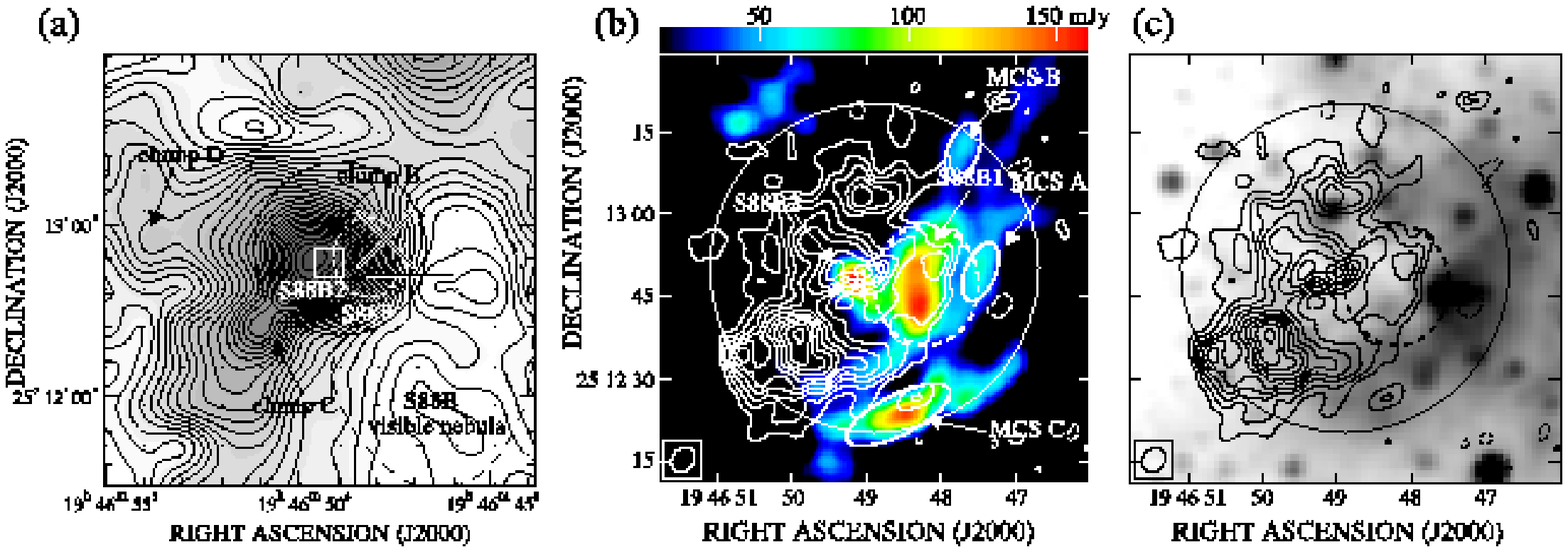}
\epsscale{0.50}
\plotone{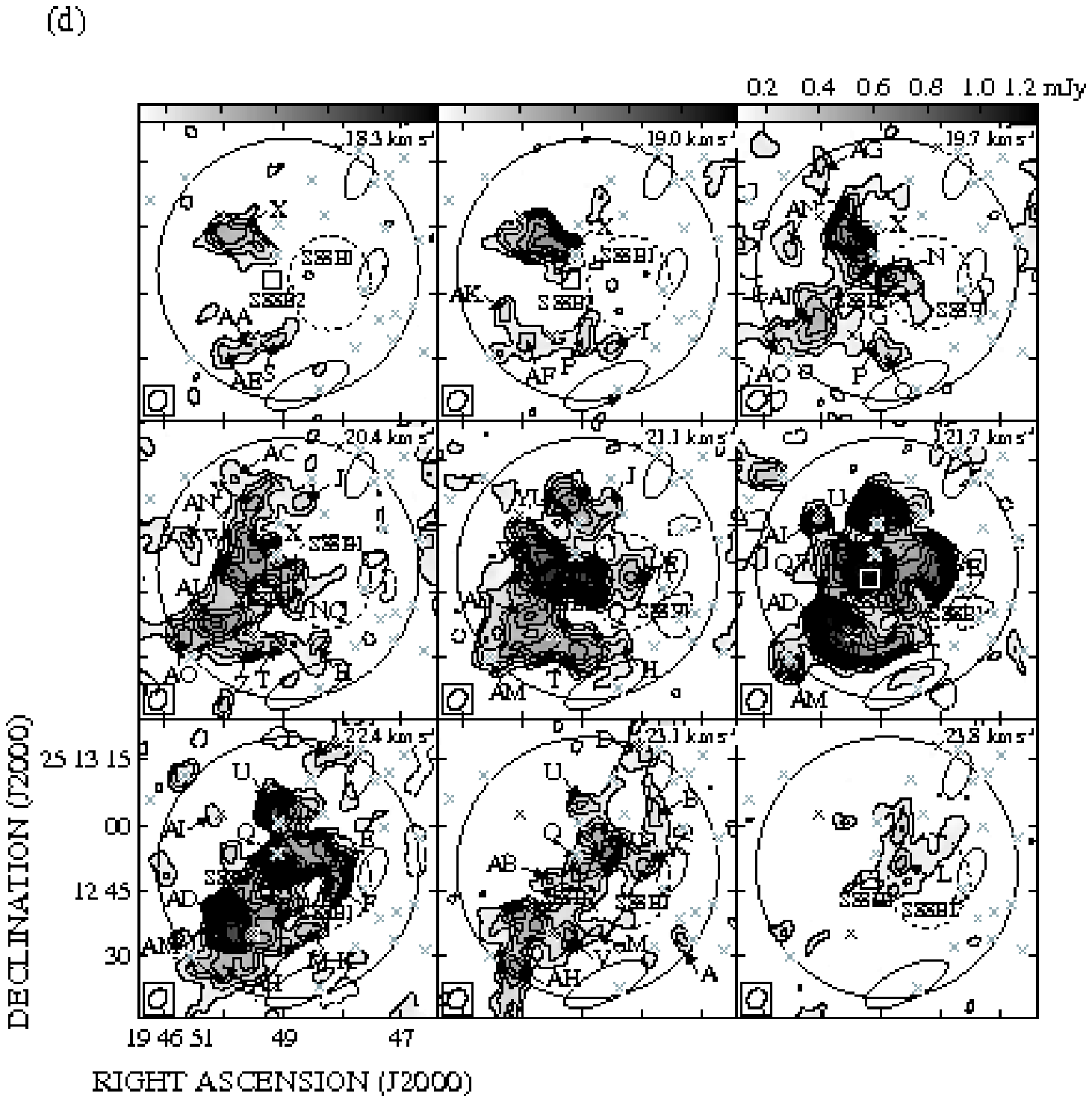}
\caption{
(a) Same as in Figure 1(a), but for IRAS\,19446+2505. The integrated velocity 
    range is between $16.5$ and $26.5$ km s$^{-1}$.
    The square and the dashed circle denote the area of the radio 
    H\,{\footnotesize II} regions S88\,B1 and B2, respectively.
    The dashed curved line denotes the boundary of the optical 
    H\,{\footnotesize II} regions S88\,B.
(b) Same as in Figure 1(b), but the color image is of the 98 GHz continuum 
    emission and the 1 $\sigma$ noise level is 13 mJy. 
(c) Same as in Figure 1(c).
(d) Same as in Figure 1(d), but the 1 $\sigma$ noise levels are 40 mJy.}
\label{fig.6}
\end{center}
\end{figure}

\clearpage 

\begin{figure}
\begin{center}
\epsscale{0.75}
\plotone{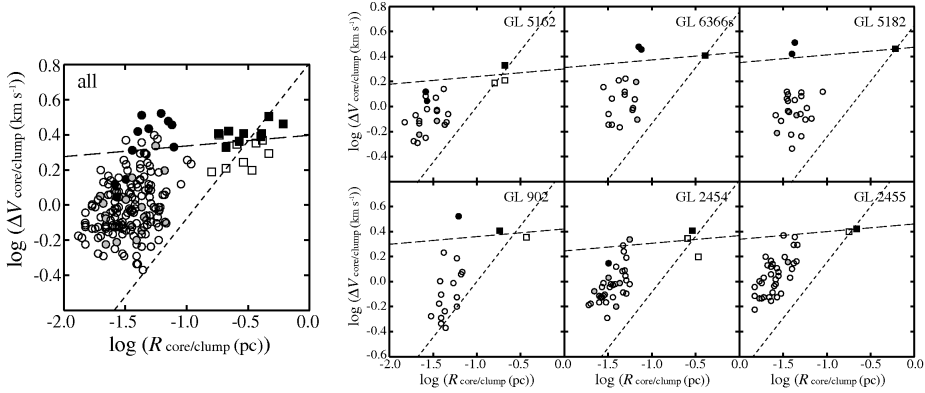}
\caption{
The correlation between the radius and the line width of the cores
and the clumps. The circles indicate the cores identified in this paper and 
by Saito et al. (2006). The squares indicate the clumps identified by 
Saito et al. (2007). 
The white, gray, and black symbols indicate the objects without 2MASS sources, 
those with 2MASS sources, and those with massive stars, respectively. 
The long- and short-dashed lines indicate the relationships of 
$\Delta V \sim R^{0.06}$ and $\Delta V \sim R^{0.89}$ using the average line
width and the average radius of the clumps in each panel, respectively}
\label{fig.7}
\end{center}
\end{figure}

\begin{figure}
\begin{center}
\epsscale{0.4}
\plotone{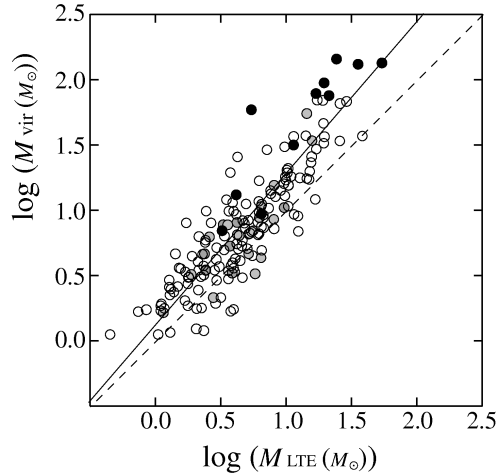}
\caption{
Virial mass plotted against the LTE mass for all cores. 
The symbols are the same as in Figure 7. The dashed line indicates 
the relationship of $M_{\rm vir}$ = $M_{\rm LTE}$. The solid line denotes 
the best fit power-law relationship for all cores except for those associated 
with massive (proto)stars, $M_{\rm vir}$ = $1.5 \times M_{\rm LTE}\,^{1.1}$.}
\label{fig.8}
\end{center}
\end{figure}

\clearpage

\begin{figure}
\begin{center}
\epsscale{0.40}
\plotone{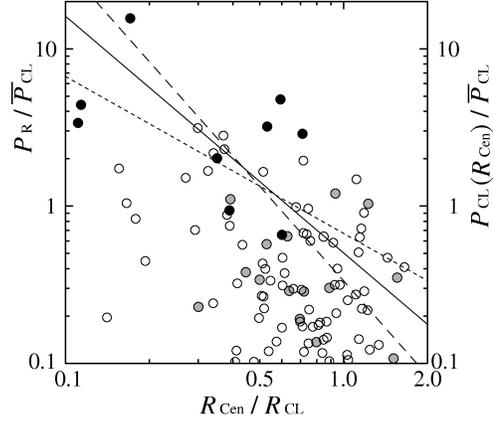}
\caption{
The ratio of the required pressure to the average clump pressure plotted 
against the distance from the center of the clump for all cores with a virial 
ratio of $> 1$. 
The dotted, solid, and dashed lines indicate the distribution of the clump 
pressure with the H$_2$ density structure of $n$(H$_2$) $\propto 
(R_{\rm Cen}/R_{\rm CL})^{-1.0}$, $\propto (R_{\rm Cen}/R_{\rm CL})^{-1.5}$, and 
$\propto (R_{\rm Cen}/R_{\rm CL})^{-2.0}$, respectively. 
The symbols are the same as in Figure 7.}\label{fig.9}
\end{center}
\end{figure}

\begin{figure}
\begin{center}
\epsscale{0.40}
\plotone{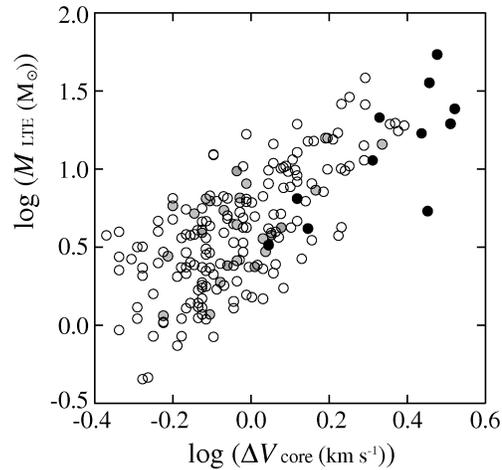}
\caption{
The LTE mass plotted against the line width of the cores. 
The symbols are the same as in Figure 7.}
\label{fig.10}
\end{center}
\end{figure}

\clearpage 

\begin{figure}
\begin{center}
\epsscale{0.35}
\plotone{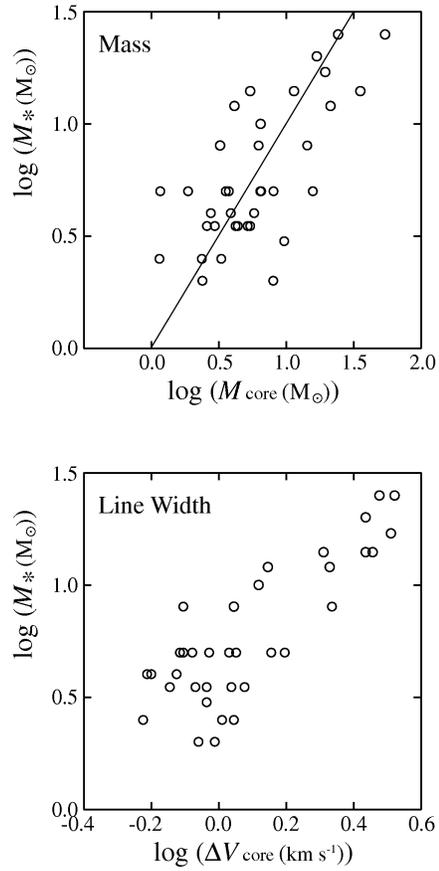}
\caption{
The mass of associated stars plotted against the LTE mass and the line width of 
the star-forming cores. The solid line indicates the relationship
of $M_{\ast}$ = $M_{\rm core}$.}
\label{fig.11}
\end{center}
\end{figure}

\clearpage 

\begin{figure}
\begin{center}
\epsscale{0.32}
\plotone{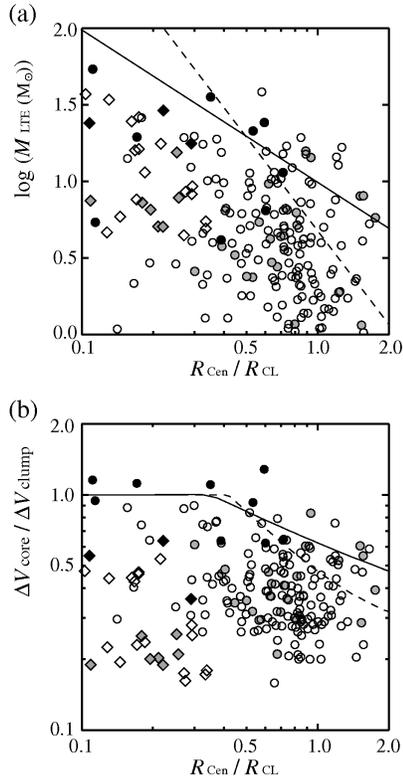}
\caption{
(a) The LTE mass of the cores plotted against the distance from the center of 
    the clump for all cores. The solid and dashed lines indicate 
    the relationship of 
    $M_{\rm core} \propto n_{0}\,(R_{\rm Cen}/R_{\rm CL})^{-\gamma}$, with 
    $\gamma$ = 1.0 and 2.0, respectively.
(b) The ratio of the line width for the cores and the clump plotted against 
    the distance from the center of the clump for all cores.
    The lines indicate the relationship between the maximum ratio of the line 
    width for the cores and the clump and the distance from the center of 
    the clump when the mass of the core has the same relationships as in 
    Figure 12(a).
    The solid and dashed lines indicate this relationship with $\gamma$ = 
    1.0 and 2.0, respectively.
The symbols are the same as in Figure 7.}
\label{fig.12}
\end{center}
\end{figure}

\clearpage 

\begin{figure}
\begin{center}
\epsscale{0.75}
\plotone{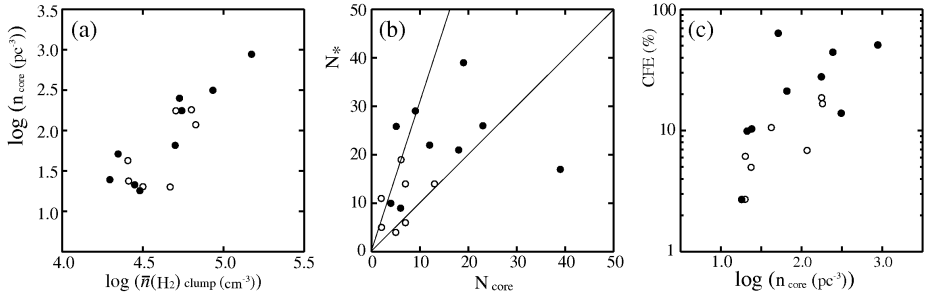}
\caption{
(a) The number density of the cores in each clump plotted against the average 
    H$_2$ density of the clump.
(b) The correlation between the number of cores and the number of associated 
    stars in each clump.
    The solid lines indicate the relationships of $N_{\ast}$ = $N_{\rm core}$ 
    and $N_{\ast}$ = $3 \times N_{\rm core}$, respectively.
(c) Correlation diagram of CFE against the number density of the cores in 
    each clump.
The open and filled circles indicate clumps without and with massive star 
formation, respectively.
}\label{fig.13}
\end{center}
\end{figure}

\clearpage 









\clearpage


\begin{table}
\begin{center}
\caption{Observation targets. \label{tab-1}}
\footnotesize
\vspace{3pt}
\begin{tabular}{cccccccccccc}
\hline
\hline
 & \multicolumn{6}{c}{Position(J2000)} & & & &\\
\cline{2-7}
\multicolumn{1}{c}{Object} & \multicolumn{3}{c}{$\alpha$} &
\multicolumn{3}{c}{$\delta$}              & Distance &    $L_{\rm IRAS} $   & 
cluster & CCS$^{a}$ & Ref.$^{b}$ \\
                           & \multicolumn{3}{c}{(h m s)}  &
\multicolumn{3}{c}{($^{\circ}$ $'$ $''$)}  & (kpc)    & (10$^{4} L_{\sun}$) &
        &              &                 \\
\hline
IRAS\,05375+3540 & 05 & 40 & 53.6 & 35 & 42 & 16 & 1.8 & \phantom{1}2.5 & 
GL\,5162  &    S235A--B   & 1 \\
IRAS\,06056+2131 & 06 & 08 & 41.0 & 21 & 31 & 01 & 3.5 & \phantom{1}7.2 & 
GL\,6366s & weak emission & 2 \\
IRAS\,06061+2151 & 06 & 09 & 07.8 & 21 & 50 & 39 & 3.5 & \phantom{1}3.4 & 
GL\,5182  & weak emission & 2 \\
IRAS\,06117+1350 & 06 & 14 & 36.6 & 13 & 49 & 35 & 3.8 & \phantom{1}6.5 & 
GL\,902   &      S269     & 3 \\
IRAS\,19442+2427 & 19 & 46 & 19.9 & 24 & 35 & 24 & 2.3 & \phantom{1}5.0 & 
GL\,2454  &       S87     & 4 \\
IRAS\,19446+2505 & 19 & 46 & 47.0 & 25 & 12 & 43 & 2.0 &           13.0 & 
GL\,2455  &   S88B1,S88B2 & 4 \\
\hline
\multicolumn{11}{@{}l@{}}{\hbox to 0pt{\parbox{180mm}{\small
\vspace{3pt}
\par\noindent
\scriptsize
$^{a}$ CCS means the centimeter continuum source. \\
$^{b}$ Data used for the distance of the target. \\
References. ---(1) Evans \& Blair 1981; (2) Moffat et al.\ 1979; 
(3) Wouterloot \& Brand 1989; (4) Crampton et al.\ 1978. \\
}\hss}}
\end{tabular}
\end{center}
\end{table}

\begin{table*}
\begin{center}
\caption{Beams and RMS noise for NMA Observations.}
\footnotesize
\vspace{3pt}
\begin{tabular}{cccccrc}
\hline
\hline
       &    \multicolumn{2}{c}{field center}     &                    &
         \multicolumn{2}{c}{beam}                   &
                           \\           
\cline{2-3}
\cline{5-6}
Object & $\alpha$(J2000) &    $\delta$(J2000)    &  Line or Continuum & 
\multicolumn{1}{c}{size} & \multicolumn{1}{c}{P.A.} & 
\multicolumn{1}{c}{r.m.s.} \\
       &      (h m s)    & ($^{\circ}$ $'$ $''$) &                    & 
($\arcsec \times \arcsec$) & \multicolumn{1}{c}{($\arcdeg$)} & 
\multicolumn{1}{c}{(mJy beam$^{-1}$)} \\
\hline
IRAS\,05375+3540 & 05 40 53.3 & 35 41 48.8 & \phantom{1}98 GHz cont. & 
$5.2 \times 4.2$ & $-$34 &            0.79  \\
                 &            &            &           110 GHz cont. & 
$4.7 \times 3.8$ & $-$36 &            0.97  \\
                 &            &            &          C$^{18}$O       & 
$4.7 \times 3.8$ & $-$65 & 110$^{a}$ \\
IRAS\,06056+2131 & 06 08 40.4 & 21 31 00.6 & \phantom{1}98 GHz cont. & 
$5.6 \times 4.4$ & $-$37 &            0.78  \\
                 &            &            &           110 GHz cont. & 
$5.0 \times 3.9$ & $-$37 &            1.08  \\
\end{tabular}
\end{center}
\end{table*}

\clearpage

\begin{deluxetable}{clcccccccc}
\tabletypesize{\scriptsize}
\tablecolumns{10}
\tablewidth{0pc}
\tablecaption{Observed Properties of C$^{18}$O cores.\label{tbl-3}}
\tablehead{
\phantom{$IRAS\,1944+2505$} & \phantom{core AA} & 
\multicolumn{2}{c}{Position$^{a}$} & & & & & & \\
\cline{3-4}
Region & core & \multicolumn{1}{c}{$\alpha$(J2000)} & 
\multicolumn{1}{c}{$\delta$(J2000)}       & \phantom{$-1$} & 
$T^{\ast}_{\rm R}\,^{b}$ & $V_{\rm LSR}\,^{b}$ & $\Delta V^{b}$ & 
$\Delta V_{\rm com}\,^{c}$ & $\int \,T^{\ast}_{\rm R}\,dV$ \\
       &      & \multicolumn{1}{c}{(h m s)}         & 
\multicolumn{1}{c}{($^{\circ}$ $'$ $''$)}  & \phantom{$-1$} & 
           (K)               &      (km\,s$^{-1}$)      &     (km\,s$^{-1}$)   &
        (km\,s$^{-1}$)     &  (K\,km\,s$^{-1}$)          
}
\startdata
IRAS\,05375+3540 & core A  & 05 40 51.3 & 35 41 40 & & 2.0 &             $-$15.9 & 0.85 & 0.92 & \phantom{$1$}1.7 \\
                 & core B  & 05 40 51.3 & 35 41 37 & & 3.0 &             $-$16.3 & 0.87 & 0.96 & \phantom{$1$}2.6 \\
                 & core C  & 05 40 52.3 & 35 41 39 & & 2.1 &             $-$18.0 & 1.09 & 1.18 & \phantom{$1$}2.3 \\
                 & core D  & 05 40 52.4 & 35 41 47 & & 2.5 &             $-$17.3 & 0.89 & 0.85 & \phantom{$1$}2.2 \\ 
                 & core E  & 05 40 52.4 & 35 42 02 & & 1.5 &             $-$18.2 & 0.51 & 0.56 & \phantom{$1$}0.8 \\
                 & core F  & 05 40 52.4 & 35 41 59 & & 1.3 &             $-$18.7 & 0.49 & 0.53 & \phantom{$1$}0.7 \\
                 & core G  & 05 40 52.8 & 35 41 18 & & 2.4 &             $-$17.1 & 0.53 & 0.60 & \phantom{$1$}1.3 \\
                 & core H  & 05 40 53.4 & 35 41 41 & & 2.3 &             $-$19.4 & 1.14 & 1.11 & \phantom{$1$}2.7 \\
\enddata
\scriptsize
\tablenotetext{a}{The position of the integrated intensity peak of C$^{18}$O cores.}
\tablenotetext{b}{The parameters are derived from a Gaussian fitting for 
the line profile at the integrated intensity peak of C$^{18}$O cores.}
\tablenotetext{c}{The line width of composite spectrum made using the data within 
the cores.}
\end{deluxetable}


\begin{table*}
\begin{center}
\caption{Observed properties of MCSs and H\,{\footnotesize II} regions.
\label{tbl-4}}
\scriptsize
\vspace{3pt}
\begin{tabular}{cccccccccc}
\hline
\hline
       & \multicolumn{2}{c}{total flux$^{a}$} &      & 
\multicolumn{2}{c}{Position$^{b}$}   & \phantom{0} & 
\multicolumn{2}{c}{total flux}                 &                  \\
\cline{2-3}
\cline{5-6}
\cline{8-9}
Region &        98\,GHz      &      110\,GHz     & name &   $\alpha$(J2000)   &     $\delta$(J2000)    & \phantom{,} & 
\multicolumn{1}{c}{98\,GHz} & \multicolumn{1}{c}{110\,GHz}    & F.F.$^{c}$ \\
       &        (mJy)        &       (mJy)       &      &        (h m s)      &  ($^{\circ}$ $'$ $''$) & \phantom{,} &  
 \multicolumn{1}{c}{(mJy)}  &   \multicolumn{1}{c}{(mJy)}     & (mJy) \\
\hline
IRAS\,05375+3540 & 263$\pm$7.0 & 258$\pm$9.2 & MCS A    & 
05 40 53.4 & 35 41 53 & & 31.9$\pm$1.4 & 46.6$\pm$1.6 & no    \\
                 &             &             & MCS B    &
05 40 53.7 & 35 41 40 & &  7.2$\pm$1.0 & 10.4$\pm$1.3 & 0.4    \\
                 &             &             & S235 A   & 
05 40 52.9 & 35 42 16 & &  108$\pm$6.0 &  105$\pm$8.2 & 250 \\
                 &             &             &          &
           &          & &              &              &         \\
IRAS\,06056+2131 & 152$\pm$4.5 & 232$\pm$6.6 & MCS A    & 
06 08 40.3 & 21 31 01 & & 41.1$\pm$1.2 & 61.6$\pm$1.7 & no  \\
\end{tabular}
\end{center}
\end{table*}

\begin{deluxetable}{cclcccccccc}
\tabletypesize{\scriptsize}
\tablecolumns{11}
\tablewidth{0pc}
\tablecaption{Derived Properties of C$^{18}$O cores.\label{tbl-3}}
\tablehead{
\phantom{$IRAS\,1944+2505$} &      & \phantom{core AA} &              & 
            &                &               &
                     &         \multicolumn{3}{c}{Associated Objects}         \\
\cline{9-11}
Region & $T_{\rm ex}$ & core & $N$(H$_{2}$)$\,^{a}$ & $R_{\rm core}$ & 
$M_{\rm core}$ & $M_{\rm vir}$ & $n$(H$_2$)$_{\rm core}$ & MCS$\,^{b}$ & 
NIR$\,^{c}$ & CCS$\,^{d}$ \\
       &     (K)     &      &  ($10^{22}$ cm$^{-2}$)  &    (pc)      & 
($M_{\sun}$)  & ($M_{\sun}$) & ($10^{5}$ cm$^{-3}$) &                   & 
                  &              
}
\startdata
IRAS\,05375+3540 & 44 & core A & \phantom{1}2.3 & 0.035 & \phantom{1}2.6 & 
\phantom{10}6.2 & \phantom{1}2.1 & ...  &  1  & ... \\
                 &    & core B & \phantom{1}3.6 & 0.027 & \phantom{1}2.9 & 
\phantom{10}5.2 & \phantom{1}5.2 & ...  & ... & ... \\
                 &    & core C & \phantom{1}3.2 & 0.034 & \phantom{1}3.7 & 
\phantom{10}9.9 & \phantom{1}3.2 & ...  & ... & ... \\
                 &    & core D & \phantom{1}3.1 & 0.020 & \phantom{1}1.8 & 
\phantom{10}3.0 & \phantom{1}7.8 & ...  & ... & ... \\
                 &    & core E & \phantom{1}1.1 & 0.026 & \phantom{1}0.9 & 
\phantom{10}1.7 & \phantom{1}1.7 & ...  & ... & ... \\
                 &    & core F & \phantom{1}0.9 & 0.019 & \phantom{1}0.5 & 
\phantom{10}1.1 & \phantom{1}2.3 & ...  & ... & ... \\
                 &    & core G & \phantom{1}1.8 & 0.022 & \phantom{1}1.2 & 
\phantom{10}1.6 & \phantom{1}3.8 & ...  &  1  & ... \\
                 &    & core H & \phantom{1}3.7 & 0.027 & \phantom{1}3.2 & 
\phantom{10}6.9 & \phantom{1}5.7 &  B   &  1  & 1 \\
\enddata
\scriptsize
\tablenotetext{a}{The parameters at the intensity peak of cores.}
\tablenotetext{b}{The name of the associated millimeter continuum source (MCS).}
\tablenotetext{c}{The number of the associated 2MASS sources.}
\tablenotetext{d}{The number of the associated centimeter continuum source.}
\tablenotetext{e}{Values cannot be derived since the radius has a large error.}
\tablenotetext{f}{This source seems to be a foreground source (see text).}
\end{deluxetable}


\begin{table*}
\begin{center}
\scriptsize
\caption{Physical properties of MCSs. \label{tbl-6}}
\vspace{3pt}
\begin{tabular}{cccccccccc}
\hline
\hline
Region & MCS & \phantom{$-$}index$^{a}$ & Type$^{b}$ & $R_{\rm d}$ & $T_{\rm d}$ &
$M_{\rm d}$   &     $n_{\rm H_2}$ & \multicolumn{2}{c}{Associated Objects} \\
       &     &       &      &     (pc)    &    (K)      & ($M_{\sun}$) & 
($10^{5}$ cm$^{-3}$) & NIR$\,^{c}$ & CCS$\,^{d}$ \\
\hline
IRAS\,05375+3540 & MCS A & \phantom{$-$}3.3$\pm$0.3 & D.T. & 0.032 & 31.5 & 
\phantom{0}35.0 & 44.7 &  1  & ...\\
                 & MCS B & \phantom{$-$}3.2$\pm$0.9 & D.T. & 0.019 & 31.5 & 
\phantom{00}9.3 & 46.1 &  1  &  1 \\
                 &       &     &      &       &      &                 &     \\
IRAS\,06056+2131 & MCS A & \phantom{$-$}3.5$\pm$0.2 & D.T. & 0.055 & 56.0 & 
111\phantom{.0} & 22.9 &  1  & ...\\
                 & MCS B & \phantom{$-$}2.2$\pm$0.3 & D.T. & 0.037 & 54.0 & 
\phantom{0}64.1 & 42.6 & ... &  2 \\
                 &       &     &      &       &      &                 &     \\
\end{tabular}
\end{center}
\end{table*}


\begin{table*}
\scriptsize
\begin{center}
\caption{Physical Properties of C$^{18}$O clumps identified by Saito et al. 
(2007). \label{tbl-7}}
\vspace{3pt}
\begin{tabular}{cccccccc}
\hline
\hline
Region & clump & $\Delta V_{\rm clump}$ & $R_{\rm clump}$ & $M_{\rm clump}$ & 
$M_{\rm vir}$ & $\overline{{\rm n(H_2)}_{\rm clump}}$ & core$^{a}$ \\
       &       &                      &     (pc)     & ($M_{\sun}$)   & 
($M_{\sun}$) & ($\times 10^{4}$ cm$^{-3}$) \\
\hline 
IRAS\,05375+3540 & clump B & 1.5 & 0.16 & \phantom{00}80 & \phantom{00}79 & 
\phantom{1}6.8 & A,B   \\
                 & clump C & 2.1 & 0.21 & \phantom{0}230 & \phantom{0}200 & 
\phantom{1}8.6 & C-M,O \\
                 & clump D & 1.6 & 0.21 & \phantom{0}170 & \phantom{0}120 & 
\phantom{1}6.4 & N,P-U \\
                 & & & & & & \\
IRAS\,06056+2131 & clump A & 2.6 & 0.41 &           1000 & \phantom{0}560 & 
\phantom{1}5.0 & A-S \\
                 & & & & & & \\
\end{tabular}
\end{center}
\end{table*}




\end{document}